\documentclass[journal ]{new-aiaa}
\usepackage[utf8]{inputenc}
\usepackage{textcomp}
\usepackage{graphicx}
\usepackage{amsmath}
\usepackage[version=4]{mhchem}
\usepackage{siunitx}
\usepackage{longtable,tabularx}
\usepackage{subfig}
\usepackage{epsfig}
\usepackage{float}
\title{Coupled Flow-Thermal Analysis of a Rocket Nozzle with Charring Ablative Thermal Protection System}

\author[1]{Basit G. Sheikh\footnote{Graduate student, Department of Aerospace Engineering, Indian Institute of Technology Kanpur, Kanpur, India.}, Rakesh Kumar\footnote{Professor, Department of Aerospace Engineering, Indian Institute of Technology Kanpur, Kanpur, India (Corresponding Author)} }
\author[2]{Susheel Kumar S\footnote{Scientist, Liquid Propulsion Systems Centre (ISRO), Thiruvananthapuram, India} }
  
\affil[1]{Indian Institute of Technology Kanpur, Kanpur, India}
\affil[2]{Liquid Propulsion Systems Centre (ISRO), Thiruvananthapuram, India}
\begin{document}

\maketitle

\begin{abstract}

This paper presents a conjugate flow–thermal analysis of a rocket nozzle protected by a charring ablative thermal protection system (TPS). The study employs a coupled approach, integrating a CFD solver with an in-house transient material response code through the exchange of boundary conditions at the fluid–solid interface. The nozzle incorporates an AVCOAT TPS and is subjected to high-temperature compressible flow. Results identify the nozzle throat as the critical location, exhibiting the highest convective loading, early attainment of the material ablation temperature, and progressive surface recession. Temporal analysis of the coupled simulations reveals an initial peak in wall heat flux followed by a transient reduction and a subsequent resurgence as viscous dissipation and evolving surface conditions modify the near-wall thermal field. At $120~s$ of simulated operation, the maximum surface recession at the throat is approximately $2.5~mm$. This research provides a methodology for predicting the thermal and ablative response of rocket nozzles equipped with charring TPS materials. The proposed framework offers valuable insights into the design and optimization of high-performance nozzles for extreme environments.
\end{abstract}

\section*{Nomenclature}

%\noindent(Nomenclature entries should have the units identified)

{\renewcommand\arraystretch{1.0}
\noindent\begin{longtable*}{@{}l @{\quad=\quad} l@{}}
$c_p$  & Specific heat capacity, J/kg K \\
$H_{\text{Pyro}}$ & Heat of pyrolysis, J/kg \\
$h$ & Convective heat transfer coefficient, W/m\(^2\) K \\
$\dot{m}_{c}^{\prime\prime}$ & Char ablation rate, kg/m\(^2\) s \\
$\dot{m}_{\text{Pyro}}^{\prime\prime}$ & Pyrolysis gas flow rate, kg/m\(^2\) s \\
$n$ & Normal coordinate \\
$P_{\text{eff}}$ & Effective thermophysical property of a material \\
$\dot{Q}_{\text{Pyro}}$ & Heat of pyrolysis per unit volume per unit time, W/m\(^3\) \\
$\dot{Q}_{\text{Tr}}$ & Transpiration cooling rate per unit volume, W/m\(^3\) \\
$Q$ & Heat of ablation (energy consumed per unit mass during ablation), J/kg \\
$\dot{q}^{\prime\prime}$ & Heat flux, W/m\(^2\) \\
$S$ & Total recession thickness, mm \\
$\dot{s}$ & Surface recession rate, mm/s \\
$T$ & Temperature, K \\
$t$ & Time, s \\
$V$ & Volume, m\(^3\) \\

\multicolumn{2}{@{}l}{\textbf{Greek Symbols}} \\
$\epsilon$ & Surface emissivity \\
$\kappa$ & Thermal conductivity, W/m K \\
$\rho$ & Density, kg/m\(^3\) \\
$\sigma$ & Stefan–Boltzmann constant, \(5.67 \times 10^{-8}\) W/m\(^2\) K\(^4\) \\

\multicolumn{2}{@{}l}{\textbf{Subscripts}} \\
blw & Blowing corrected \\
c & Char \\
conv & Convective \\
g & Gas \\
r & Recovery \\
v & Virgin \\
w & Wall \\
$\infty$ & Freestream \\
\end{longtable*}}

\section{Introduction}

\lettrine{R}{ocket} nozzle, a critical component of any propulsion system, must endure the extreme thermal loads during its operation. The high temperature and high speed gases exiting the combustion chamber interact with the nozzle wall and potentially degrade the nozzle material. In order to maintain and safeguard the structural integrity, a nozzle is incorporated with a thermal protection system (TPS). TPS materials, such as graphite, carbon-carbon, carbon-phenolic, silica-phenolic, and filled rubbers are particularly designed to sustain such extreme heat loads. These materials may undergo mass loss (ablation) when exposed to high-temperature environments in order to protect the underlying nozzle structure. The primary objective of a TPS is to shield the nozzle surface from the intense heat. Hence, the design of a TPS demands a comprehensive understanding of interactions between high temperature gas flow and TPS material. For the optimal performance of a rocket nozzle, it is essential to obtain an accurate thermal response of a TPS. While maximizing the TPS thickness may enhance the protection, but can put weight and space constraints. Conversely, minimizing the TPS thickness may compromise the protection, leading to system failure. A typical factor of safety used in the TPS design demands initial thickness to be at least twice the anticipated surface recession and 1.25 times the projected final char layer thickness \cite{powers1981shuttle,arnold1979subscale}. Therefore, designing a TPS that ensures the safety but maintaining minimal weight and space requirements is crucial. 

Several studies to understand the behavior of gas flows through nozzles demonstrate that variations in nozzle geometry and operating conditions hugely influence its performance \cite{darbandi2011study,stark2013flow,rakhsha2023effect,lijo2010numerical,schneider2018numerical,balabel2011assessment,sathish2017heat}. Research on nozzle surface ablation is broadly divided into two categories: experimental investigations \cite{thongsri2022gas,sae2021insulation,hui2017ablation} and computer simulations \cite{cross2018reduced,zhang2022numerical,babu2020prediction}. Experiments conducted by Sae-ngow \textit{et al.}\cite{sae2021insulation} analyzing ablation in static supersonic nozzles focused on silica phenolic composites. Their findings highlight that silica phenolic materials outperform rubber based thermal insulators under same operating conditions. Hui \textit{et al.} \cite{hui2017ablation} also performed experiments on a de Laval nozzle to investigate the relationship between nozzle geometry, gas flow behavior and ablation patterns. Their study revealed that the maximum ablation occurs at the throat of the nozzle. Despite experimental results providing valuable insights and being reliable, they are costly and time-consuming. These limitations have made researchers shift their focus on the computational studies.  

Traditionally, thermal behavior and ablation analysis was decoupled, where both the flow solver and the material response solver were handled independently. Early methods like those proposed by Bartz \cite{bartz1957simple} used simple correlations to calculate the convective heat flux. However, literature has shown that the material response of the nozzle is influenced by the surface temperature profile considered while calculating the wall heat flux \cite{cross2017conjugate}. As a result,  such decoupled methods lacked fidelity to capture the intricate interactions between flow field and the solid domain.

A more accurate and robust approach to study the ablation of rocket nozzles is by employing conjugate heat transfer (CHT) analysis \cite{cross2019conjugate,zhang2011coupled,ding2017transient,pizzarelli2016validation,guan2017conjugate}. In such analysis, the flow solver is coupled with a material thermal response solver. The coupling of these solvers include sharing the boundary properties like heat flux and surface temperatures at the fluid-solid interface. Additionally, to account for the surface recession and the geometry changes due to ablation, the coupling between flow and thermal solver becomes essential \cite{kuntz2001predictions}. 

The initial effort to perform a coupled analysis of flow field and ablation was done by NASA focusing on a pyrolyzing TPS material \cite{olynick1999aerothermodynamics}. This study examined transient, one-dimensional ablation in the Stardust sample return capsule. However, the investigation did not account for surface recession due to ablation. Kuntz \textit{et al.} at Sandia National Laboratories \cite{kuntz2001predictions} performed a CHT analysis for the nose tip of  IRV-2 vehicle, by integrating transient, two-dimensional ablation. However, this study had a limitation that they considered only non-charring TPS materials. Researchers at NASA researchers also analyzed the same test case using a similar approach \cite{thompson2008implementation}, but their analysis was restricted to a one-dimensional ablation assumption.

Researchers at the University of Michigan \cite{wiebenga2012computation,alkandry2013coupled, doi:10.2514/1.A32847} also conducted ablation analyses for IRV-2 vehicle and the Stardust reentry capsule using transient response from the material. However, their study was also limited to  non-charring materials. Similarly, Thakre and Yang \cite{thakre2008chemical} investigated nozzle flows under the steady-state assumption focusing on non-charring materials with one-dimensional ablation. Cross and Boyd \cite{cross2019conjugate} performed ablation analysis for HIPPO nozzle, where they coupled the LeMANS solver and MOPAR-MD material solver. They used an algorithm exclusively developed and utilized by their team, which produced reliable ablation results. Since the methodology and code used were not made public, it is difficult for other researchers to advance and build upon their work. 
 
The primary goal of the present study is to perform conjugate thermal analysis by loosely coupling a commercial CFD solver (ANSYS fluent) with an in-house material response code, charring ablator thermal solver (CATS). The surface temperature distribution obtained from the thermal response solver is transferred to the flow solver that gives surface heat flux to the thermal response solver, thereby enabling the interaction between the two solvers. A user defined function is utilized to apply the non-uniform temperature distribution on the nozzle wall in order to couple CFD solver with thermal solver. The study presents the temperature and heat flux distribution, and the resulting surface thermal ablation for a charring material.

The organization of the paper is as follows: the paper begins with a brief introduction of flow and material solvers, followed by a description of the flow-thermal coupling approach. A case from a technical NASA report \cite{back1964convective} is used to validate the flow solver. After validation, the results for flow through a rocket nozzle are discussed with a primary focus on heat flux calculations and the resulting temperature rise within the TPS. Finally the paper concludes with a summary of key findings, highlighting the significance of the results.

\section{Methodology}

\subsection{Governing Equations for Flow Solver \cite{manual2009ansys}}

For the nozzle flow problem, the governing equations consist of the continuity, momentum, and energy equations.
\textbf{Continuity equation} ensures mass conservation through
\begin{equation}
    \frac{\partial \rho}{\partial t} + \nabla \cdot (\rho \vec{V}) = 0,
\end{equation}
where \(\rho\) is density and \(\vec{V}\) represents velocity.
 
For \textbf{momentum conservation} in each spatial direction \(x, y, z\), the equation
\begin{equation}
    \frac{\partial (\rho \vec{V})}{\partial t} + \nabla \cdot (\rho \vec{V} \vec{V}) = -\nabla p + \nabla \cdot \tau + \rho \vec{g}
\end{equation}
holds, incorporating pressure \(p\), stress tensor \(\tau\), and gravitational force \(\vec{g}\).
 
\textbf{Energy conservation} is modeled through
\begin{equation}
    \frac{\partial (\rho E)}{\partial t} + \nabla \cdot \left( \vec{V}(\rho E + p) \right) = \nabla \cdot \left( k \nabla T \right) + \Phi,
\end{equation}
where \(E\) denotes total energy, \(T\) temperature, \(k\) thermal conductivity, and \(\Phi\) is the viscous dissipation term.

To accurately simulate turbulent flow in a nozzle, three turbulence models were utilized, viz., Spalart-Allmaras model, the Baseline \(k\)-\(\omega\) model and the Shear Stress Transport (SST) \(k\)-\(\omega\) model.

The Spalart-Allmaras (SA) model solves a single equation for turbulent viscosity:
\begin{equation}
    \frac{\partial (\rho \tilde{\nu})}{\partial t} + \nabla \cdot (\rho \tilde{\nu} \vec{V}) = G_{\tilde{\nu}} - Y_{\tilde{\nu}} + S_{\tilde{\nu}},
\end{equation}
where \(\tilde{\nu}\) represents modified turbulent viscosity.

The Baseline (BSL) \(k\)-\(\omega\) model, combining the \(k\)-\(\omega\) formulation for boundary layers with \(k\)-\(\epsilon\) for the free stream, and and the Shear Stress Transport (SST) \(k\)-\(\omega\) model, which blends boundary-layer sensitivity and free-stream behavior for better precision, solve transport equations for turbulent kinetic energy and specific dissipation:
\begin{equation}
    \frac{\partial (\rho k)}{\partial t} + \nabla \cdot (\rho k \vec{V}) = P_k - \beta^* \rho k \omega + \nabla \cdot \left( \frac{\mu_t}{\sigma_k} \nabla k \right),
\end{equation}
\begin{equation}
    \frac{\partial (\rho \omega)}{\partial t} + \nabla \cdot (\rho \omega \vec{V}) = \alpha \frac{\omega}{k} P_k - \beta \rho \omega^2 + \nabla \cdot \left( \frac{\mu_t}{\sigma_\omega} \nabla \omega \right),
\end{equation}

Among these, the BSL \(k\)-\(\omega\) model emerged as the most accurate for capturing boundary-layer separation and heat transfer in high-speed nozzle flows, a crucial aspect for accurate predictions.

\subsection{Governing Equations for Material Response Solver}

Charring ablative materials are widely used in TPS for high-speed re-entry vehicles, owing to their ability to dissipate heat by sacrificially losing mass through pyrolysis and ablation. The CATS \cite{appar2022conjugate} was utilized in this study to model this thermal response, accounting for heat conduction, pyrolysis, transpiration cooling, and surface recession.

The primary governing equation for heat conduction within charring ablative materials is expressed as
\begin{equation}
    \int_V \left( \rho c_p \frac{\partial T}{\partial t} \right) dV = \int_A \kappa(T) \frac{\partial T}{\partial n} dA - \int_V \left( \dot{Q}_{\text{pyro}} + \dot{Q}_{\text{trans}} \right) dV,
\end{equation}
where \(T\) is the temperature, \(\rho\) is the material density, \(c_p\) is the specific heat capacity, \(\kappa(T)\) is the temperature-dependent thermal conductivity, and \(\dot{Q}_{\text{pyro}}\) and \(\dot{Q}_{\text{trans}}\) represent the heat absorbed by pyrolysis and transpiration cooling, respectively. The thermal boundary conditions are defined by heat exchange at the exposed surface, where
\begin{equation}
    \kappa(T) \frac{\partial T}{\partial n} \bigg|_{\text{surface}} = \dot{q}_{\text{conv}} - \epsilon \sigma (T_w^4 - T_\infty^4) - \dot{m}_c Q^*,
\end{equation}
with \(\dot{q}_{\text{conv}}\) representing convective heat flux, \(\epsilon\) emissivity, \(\sigma\) the Stefan-Boltzmann constant, \(T_w\) the wall temperature, \(T_\infty\) the surrounding temperature, \(\dot{m}_c\) the mass loss rate per unit area, and \(Q^*\) the heat of ablation. The back surface is assumed insulated:
\begin{equation}
    \kappa(T) \frac{\partial T}{\partial n} \bigg|_{\text{back}} = 0.
\end{equation}
The pyrolysis process, an endothermic reaction that decomposes material and releases pyrolysis gases, absorbs heat per unit volume as follows:
\begin{equation}
    \dot{Q}_{\text{pyro}} = \dot{\rho} H_{\text{pyro}},
\end{equation}
with \(H_{\text{pyro}}\) denoting the heat of pyrolysis. Pyrolysis gases cool the char layer through transpiration cooling, modeled as
\begin{equation}
    \dot{Q}_{\text{trans}} = \dot{\rho} c_{p,g} (T - T_g),
\end{equation}
where \(c_{p,g}\) is the specific heat of the gas and \(T_g\) is the gas temperature. When the material surface reaches ablation temperature \(T_{\text{abl}}\), surface recession begins, characterized by
\begin{equation}
    \dot{S} = \frac{\dot{m}_c}{\rho_{\text{char}}},
\label{eq:surface recession}
\end{equation}
with \(\dot{S}\) as the recession rate, \(\dot{m}_c\) the mass loss rate, and \(\rho_{\text{char}}\) the char density.

Pyrolysis gas provides a major contribution in lowering the heat load on the surface of charring ablative TPS materials. This gas is a by-product of pyrolysis process and helps in heat removal from the material via transpiration cooling. It forms a thin protective layer between the incoming high temperature gases and the material. This study does not explicitly model the physical ejection of pyrolysis gases from the material surface. However, the impact of pyrolysis gas blowing on heat flux predictions is incorporated using a correlation function developed by Kays~\cite{kays1980convective} and applied in recent research by Phadnis {\em et al.}~\cite{phadnis2020effect}. The governing relationship is expressed as:

\begin{equation}
    \dot{q}''_{\text{blw}} = \Omega \dot{q}''_{\text{conv}},
    \label{eq:blowing_correction}
\end{equation}

where the dimensionless blowing correction parameter \( \Omega \) is calculated as:

\begin{equation}
   \Omega = \frac{\psi}{e^\psi - 1},
    \label{eq:X_parameter}
\end{equation}

with \( \psi \) being a non-dimensionalized function of the mass flux from the material into the boundary layer, given by:

\begin{equation}
    \psi = \frac{c_{p,g} \dot{m}''_g}{h},
    \label{eq:w_parameter}
\end{equation}

and the convective heat transfer coefficient \( h \) is defined as:

\begin{equation}
    h = \frac{\dot{q}''_{\text{conv}}}{T_r - T_w}.
    \label{eq:heat_transfer_coefficient}
\end{equation}

Here, \( \dot{q}''_{\text{conv}} \) is the convective heat flux, \( T_r \) is the recovery boundary layer temperature, and \( T_w \) is the wall temperature.

The blowing correction functions for heat flux, as expressed in Eqs.~\eqref{eq:blowing_correction}--\eqref{eq:heat_transfer_coefficient}, are directly implemented in the thermal solver to account for the heat flux reduction due to pyrolysis gas blowing at the surface boundary. This approach ensures an accurate representation of the thermal response of the TPS material under extreme heat loads.

\subsection{Flow-Thermal Coupling}

In conjugate heat transfer of nozzle flows, the interaction between the hot gases and the material of the nozzle is referred to as flow-thermal coupling. In this work, it essentially involves the exchange of wall boundary conditions between the flow and solid domain of the nozzle. In order to achieve this, the commercial CFD flow solver, (ANSYS Fluent)\cite{manual2009ansys}, is coupled to an in-house material thermal response solver, Charring Ablator Thermal Solver (CATS). The coupling strategy enables the conjugate simulations of ablation of a charring and pyrolyzing material\cite{doi:10.1080/10618562.2021.2017900}.

In literature, the flow-thermal solvers are coupled using different strategies.
For instance \cite{appar2022conjugate} has coupled the flow and thermal solvers using two different ways: non-iterative and iterative coupled approaches. However this study was done for hypersonic reentry vehicles. The coupling strategy utilized in this study is based on non-iterative approach. This approach was seen to be computationally efficient as compared to the other method. In this method, the two solvers exchange the boundary conditions at the fluid-solid interface.

The interaction between flow and thermal response solver is outlined as follows:
\begin{enumerate}
    \item A steady-state flow solution is obtained at the initial time point $t = t_0$, using ANSYS Fluent, with an isothermal boundary condition applied at the solid surface.
    \item The solution of flow solver (convective heat flux distribution along the nozzle wall) is then passed to the in-house thermal response solver, CATS.  
    \item The thermal response code solves the transient heat conduction equation within the nozzle wall material, advancing in time from $t_0$ to $t_0 + \Delta t$, with keeping the input heat flux constant. This captures the nozzle’s response over a small operational interval.
    \item As the thermal response solver marches forward in time, at the new time point $t = t_1$, the updated wall temperature $T_w$ is transferred as a Dirichlet boundary condition to the flow solver.
    \item Finally, using the updated surface temperature, the flow solver then computes a new steady-state convective heat flux distribution, thus providing input for the next thermal coupling step.
\end{enumerate}

This entire process of coupling the properties at the fluid-solid interface relies on the coupling time-steps being very small. This ensures that the assumption of constant heat flux between two time steps is within tolerance limit.

\section{Validation Case}\label{Validation}

In this study, \textsc{ANSYS Fluent} is employed to perform the CFD simulations, while an in-house thermal response solver is used to predict the material response of the TPS. The latter has already been comprehensively validated in previous works~\cite{appar2022conjugate,doi:10.1080/10618562.2021.2017900}. 

\begin{figure}[hbt!]
\centering
\includegraphics[width=1\textwidth]{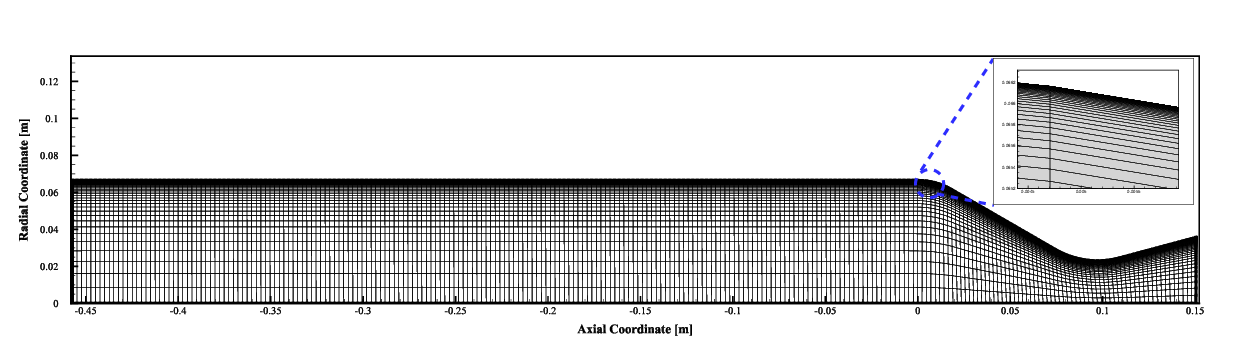}
\caption{Nozzle with cooling duct mesh}
\label{fig:Nozzle with cooling duct mesh.}
\end{figure}

To validate the flow solver (\textsc{ANSYS Fluent}), a benchmark test case from the NASA Technical Report by Back \emph{et al.}~\cite{back1964convective} is considered. Specifically, Case 262 is reproduced, which involves supersonic flow through a convergent--divergent nozzle with wall cooling. The configuration provides detailed experimental data on wall heat transfer and pressure distribution, making it a standard validation case for nozzle heat-transfer studies. The nozzle geometry and operating conditions are replicated from the experimental setup, and the computational mesh of the flow domain is shown in Fig.~\ref{fig:Nozzle with cooling duct mesh.}. Details of the nozzle geometry are provided in Table~\ref{tab:nozzle_geometry}.

\begin{figure}[hbt!]
    \centering
    \subfloat[Heat transfer coefficient for different grid resolutions.]{
        \includegraphics[width=0.48\textwidth]{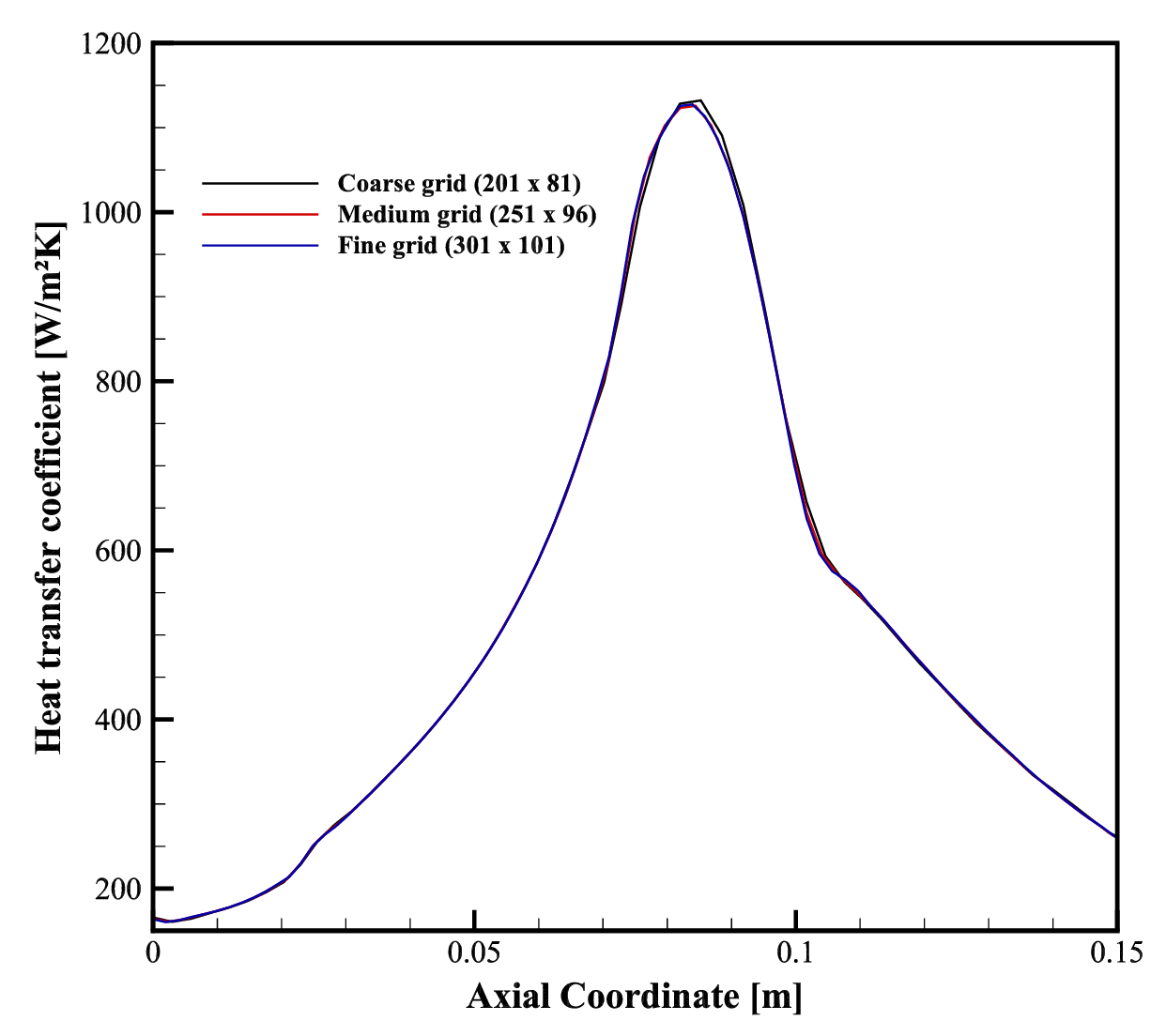}
        \label{fig:GridIndependence}
    }
    \hfill
    \subfloat[\(Y^+\) distribution along the nozzle wall for the medium grid.]{
        \includegraphics[width=0.48\textwidth]{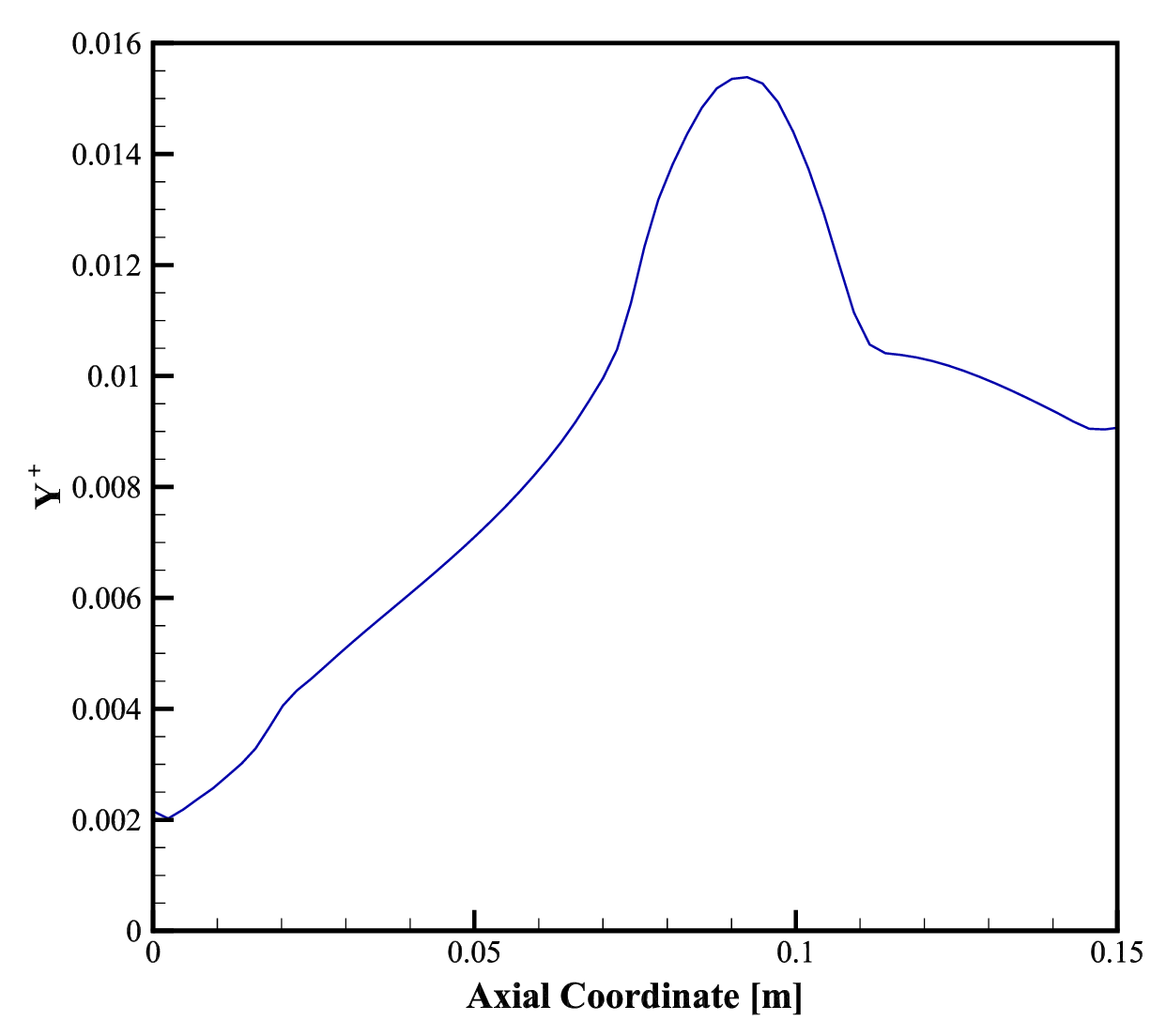}
        \label{fig:YPlus}
    }
    \caption{Grid independence and near-wall resolution assessment for the nozzle flow simulations.}
\end{figure}

\begin{table}[h!]
    \centering
    \begin{minipage}{0.49\textwidth}
        \centering
        \caption{Nozzle Geometry Specifications}
        \label{tab:nozzle_geometry}
        \begin{tabular}{|l|c|}
        \hline
        \textbf{Parameter}                    & \textbf{Value}       \\ \hline
        Throat diameter                       & 0.0458 m             \\ \hline
        Contraction area ratio (AR)           & 7.75:1               \\ \hline
        Expansion area ratio (AR)             & 2.68:1               \\ \hline
        Convergent half-angle                 & 30°                  \\ \hline
        Divergent half-angle                  & 15°                  \\ \hline
        \end{tabular}
    \end{minipage}
\hfill
    \begin{minipage}{0.49\textwidth}
        \centering
        \caption{Operating Conditions}
        \label{tab:operating_conditions}
        \begin{tabular}{|l|c|}
        \hline
        \textbf{Condition}                    & \textbf{Value}       \\ \hline
        Nozzle inlet pressure                 & $5.171 \times 10^5$ N/m\textsuperscript{2} \\ \hline
        Nozzle inlet temperature              & 843.33 K             \\ \hline
        Nozzle inlet density                  & 2.13 kg/m\textsuperscript{3} \\ \hline
        Transport model                       & Sutherland’s law     \\ \hline
        Turbulence model                      & SA, BSL, SST \\ \hline
        Exit Mach Number                      & 2.5 \\ \hline
        \end{tabular}
    \end{minipage}
\end{table}
The grid independence study was conducted to verify that the computed results were not influenced by mesh resolution. Three grids: coarse (\(201 \times 81\)), medium (\(251 \times 96\)), and fine (\(301 \times 101\)) were tested. Figure~\ref{fig:GridIndependence} shows the variation of the heat transfer coefficient, where minimal differences were observed between the medium and fine grids. To accurately capture wall properties, the first layer of cells was positioned close to the wall to ensure adequate resolution of the boundary layer. Additionally, the mesh was refined near the nozzle wall to capture heat transfer phenomena effectively, maintaining the Y$^+$ value way below 1 throughout the nozzle wall. Figure~\ref{fig:YPlus} confirms that the \(Y^+\) values remained below 1 for all grids, ensuring accurate resolution of the viscous sublayer. A structured mesh with 251 $\times$ 96 elements was used for our simulations. 

 The test case parameters and boundary conditions are summarized in Table~\ref{tab:operating_conditions}. An isothermal boundary condition was applied to the nozzle wall, with the wall temperature set to half of the nozzle inlet temperature.

\begin{figure}[hbt!]
\centering
\includegraphics[width=0.6\textwidth]{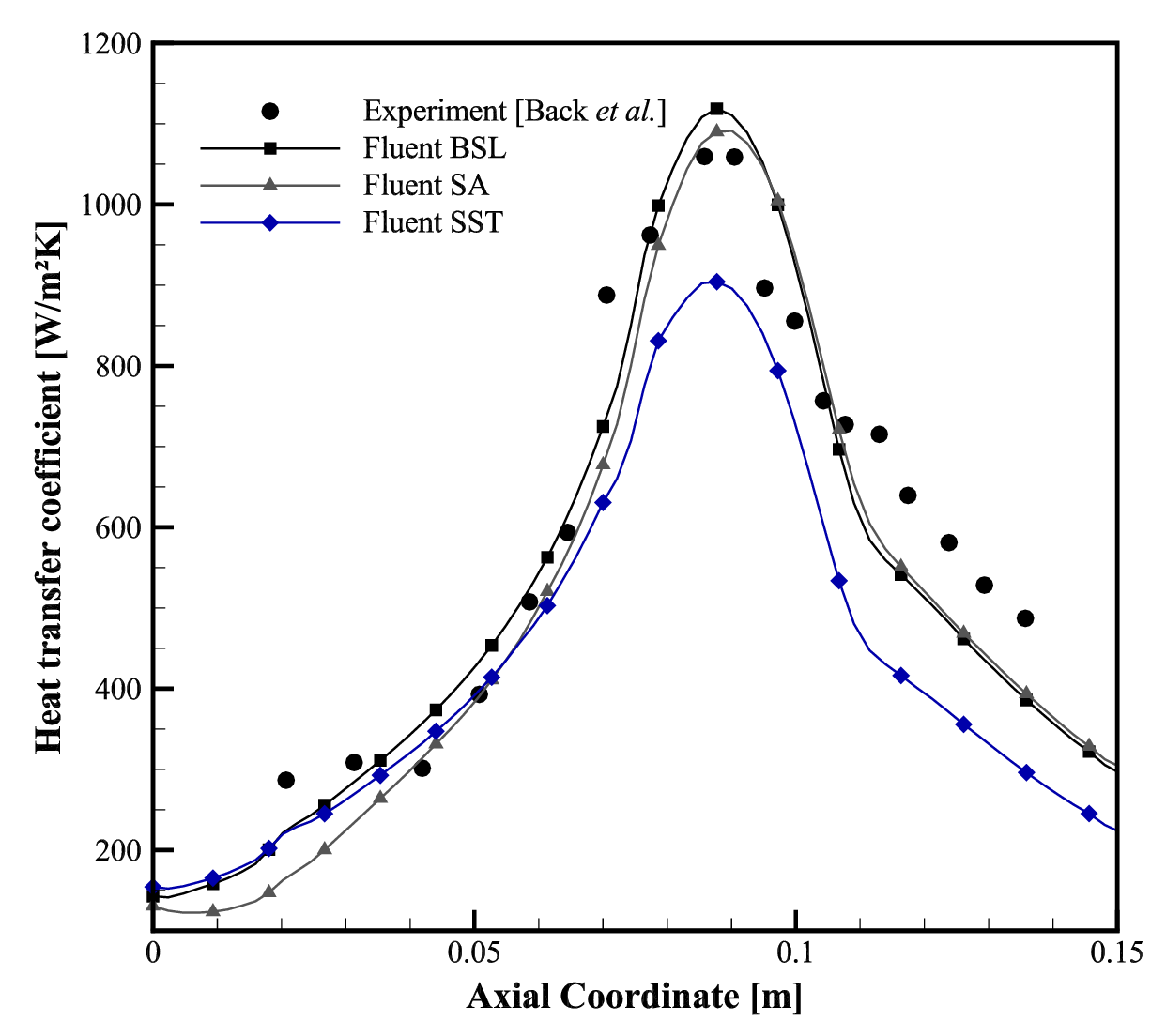}
\caption{Heat transfer coefficient using different turbulence models}
\label{fig:Validation}
\end{figure}

To validate the heat transfer coefficient variation along the wall, three different turbulence models are used: the Spalart-Allmaras (SA), Baseline (BSL) \(k\)-\(\omega\), and Shear Stress Transport (SST) \(k\)-\(\omega\) models. The results are provided in Fig. \ref{fig:Validation}, wherein it can be noticed that the BSL and SA model show a good agreement with the experimental data.

\begin{figure}[hbt!]
    \centering
    \subfloat[Mach number contour]{
        \includegraphics[width=0.48\textwidth]{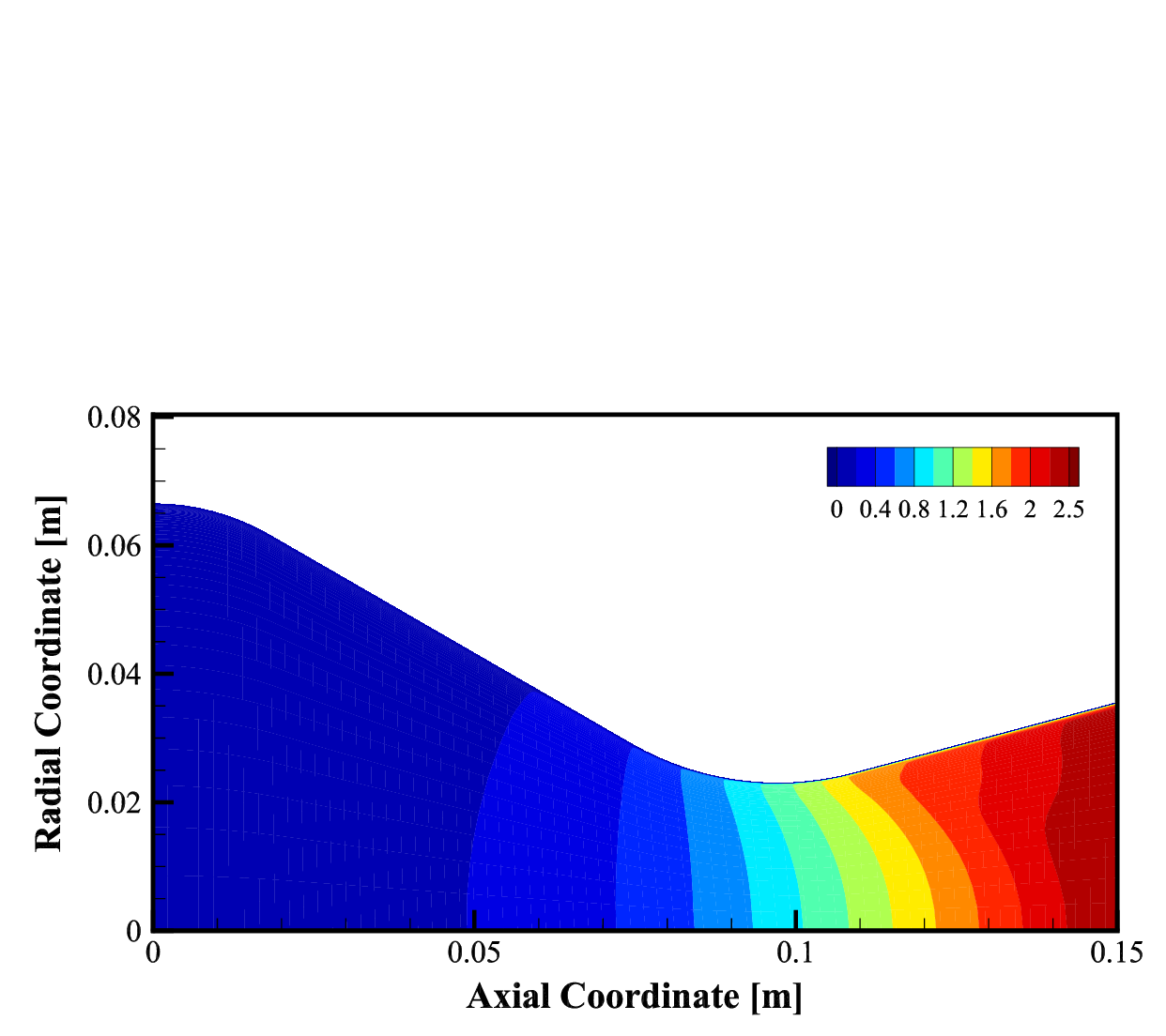}
        \label{fig:Mach Contour.}
    }
    \hfill
    \subfloat[Mach number variation along the axis]{
        \includegraphics[width=0.48\textwidth]{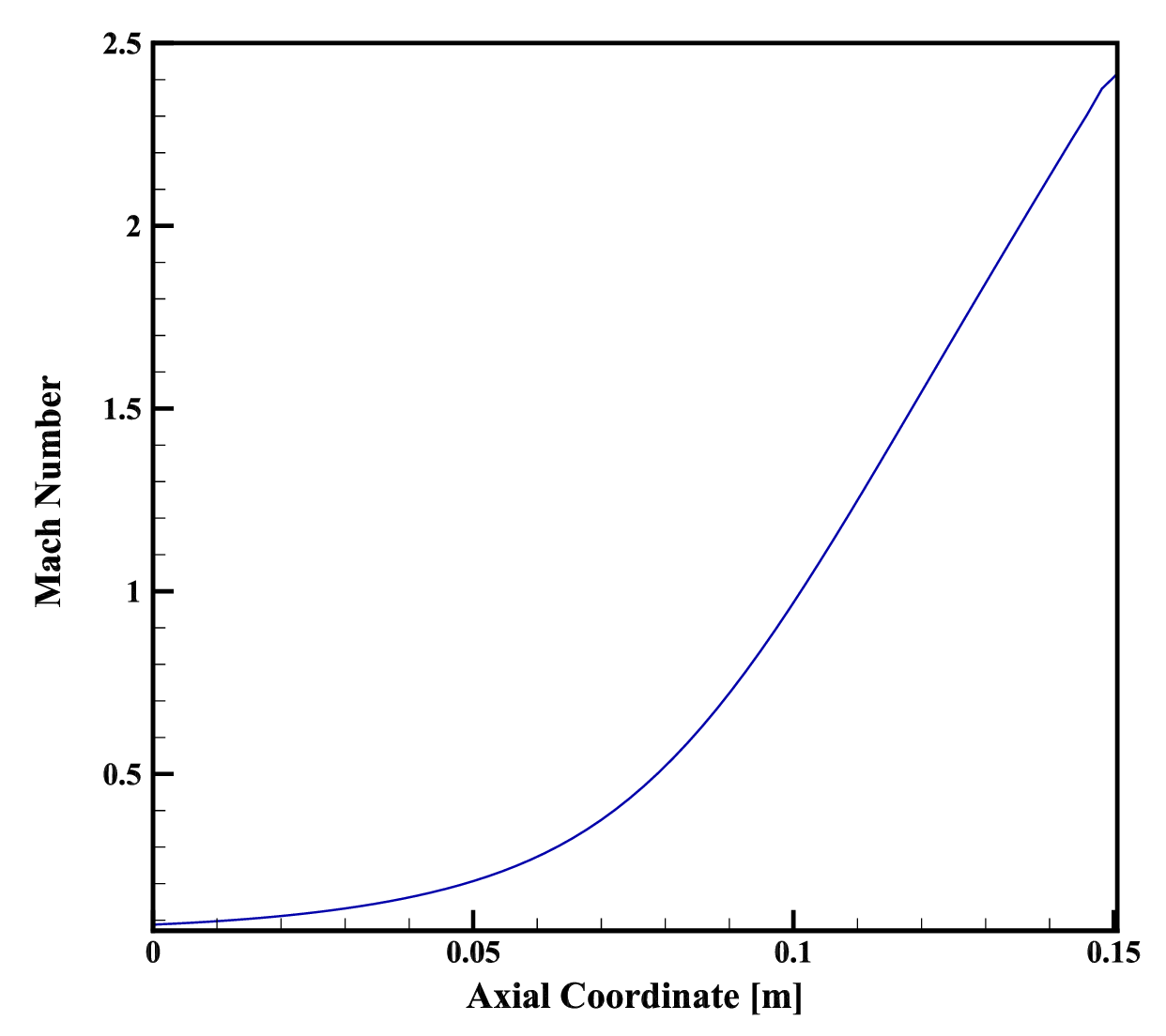}
        \label{fig:Mach Number Along the axis}
    }
    \caption{Mach Number Variation.}
\end{figure}

\begin{figure}[hbt!]
    \centering
    \subfloat[Temperature contour]{
        \includegraphics[width=0.48\textwidth]{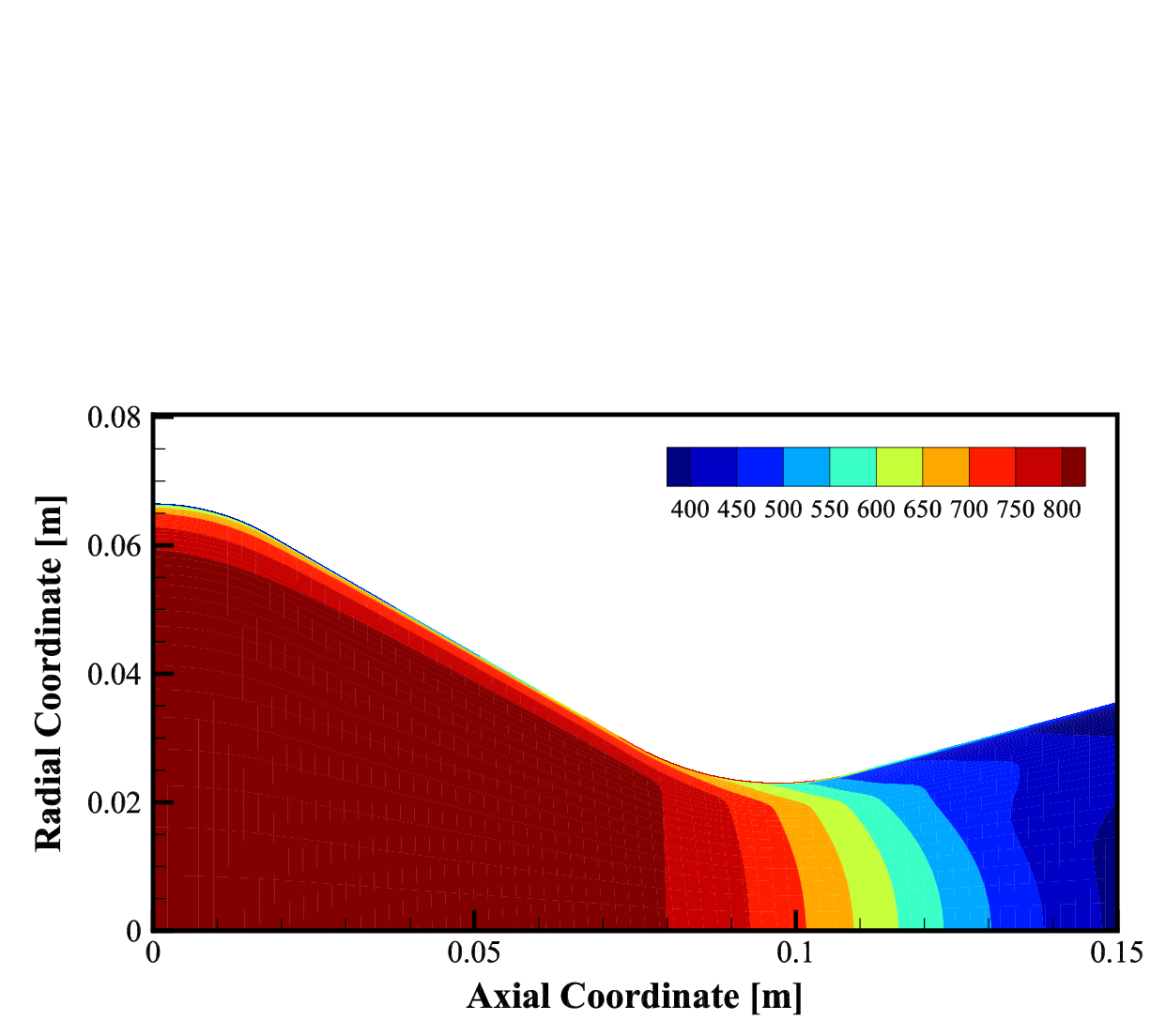}
        \label{fig:Temperature Contour}
    }
    \hfill
    \subfloat[Temperature variation along the axis]{
        \includegraphics[width=0.48\textwidth]{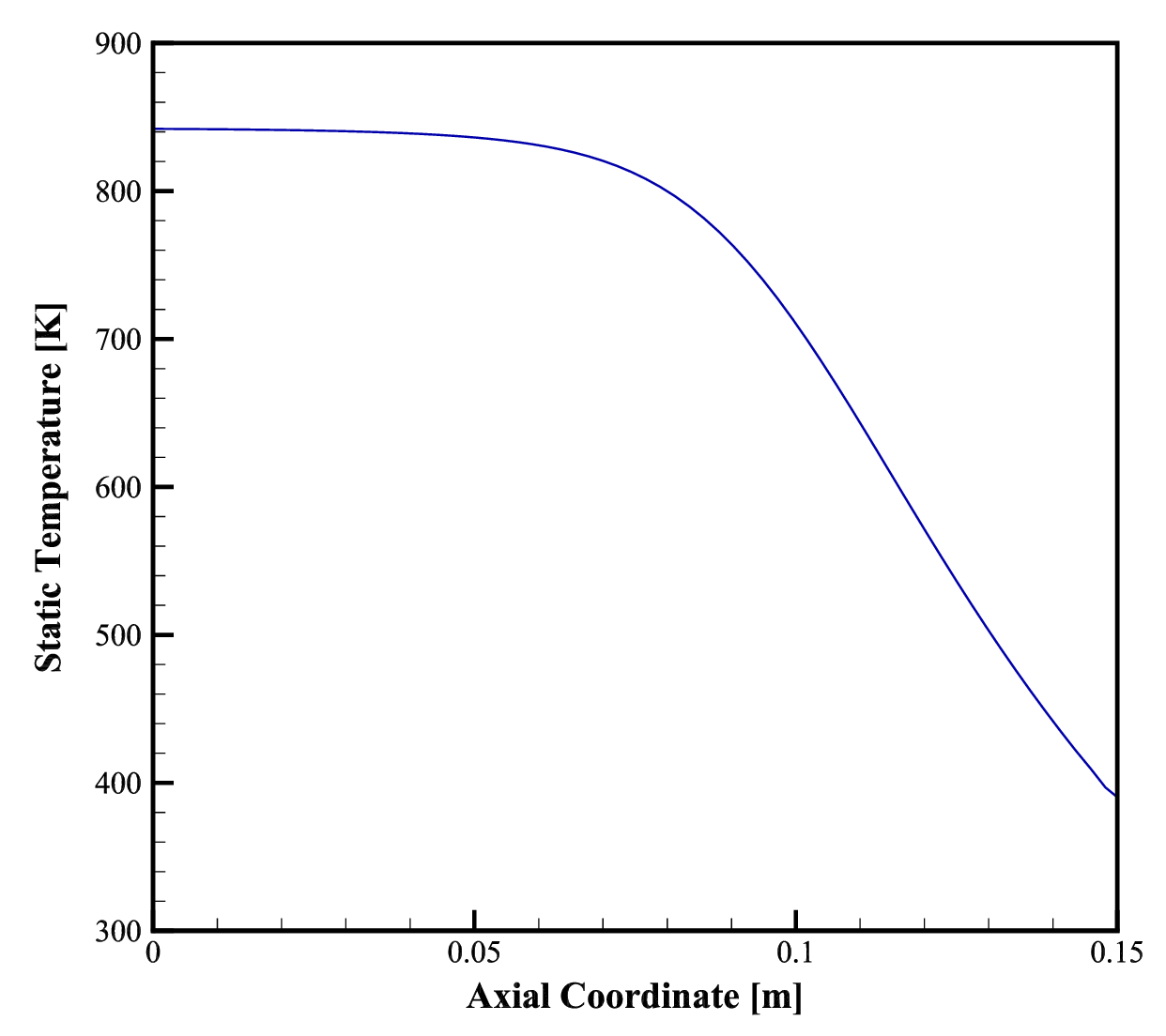}
        \label{fig:Temperature variation along the axis}
    }
    \caption{Temperature variation}
\end{figure}

\begin{figure}[hbt!]
    \centering
    \subfloat[Pressure contour]{
        \includegraphics[width=0.48\textwidth]{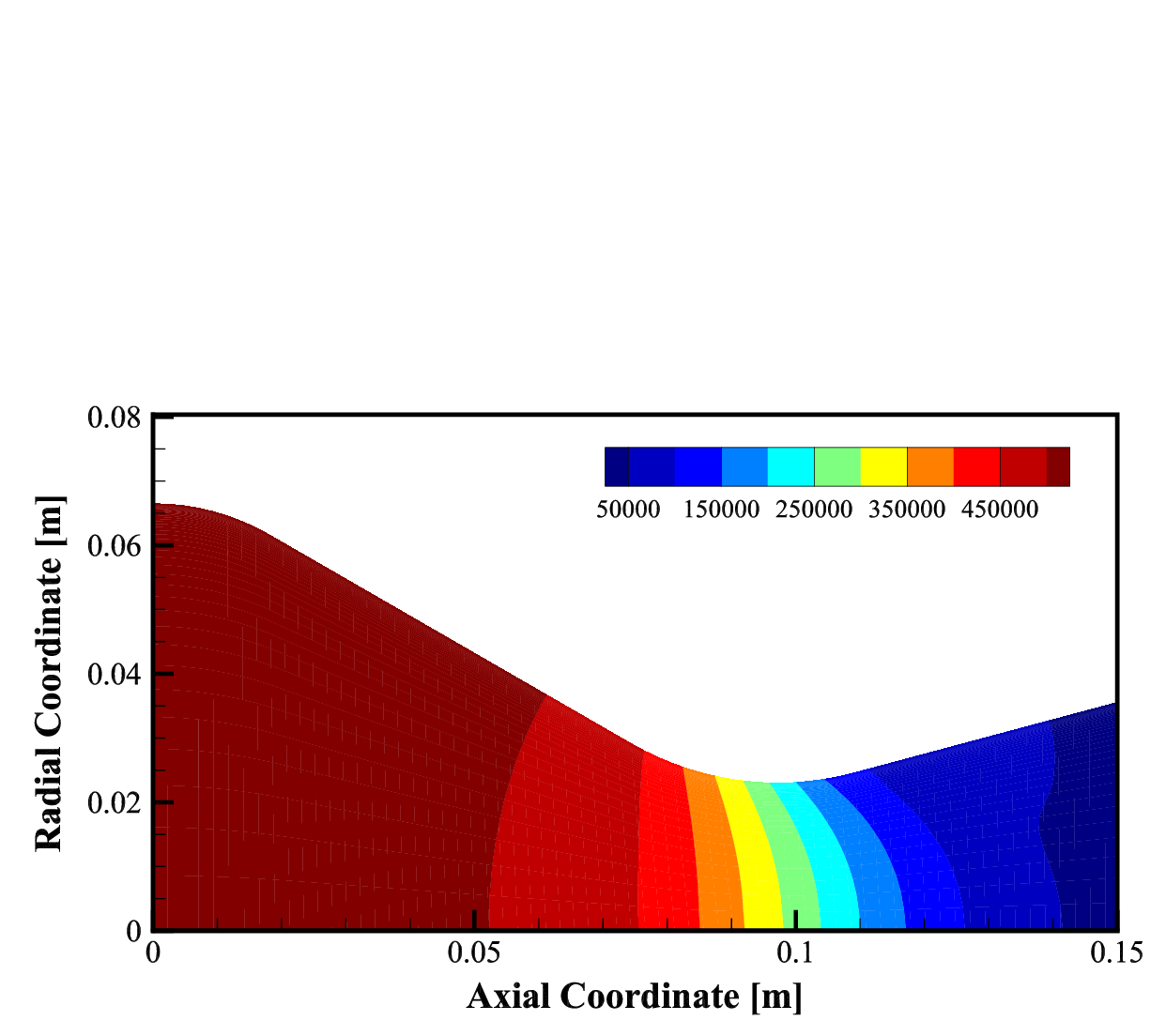}
        \label{fig:Pressure Contour}
    }
    \hfill
    \subfloat[Pressure variation along the axis]{
        \includegraphics[width=0.48\textwidth]{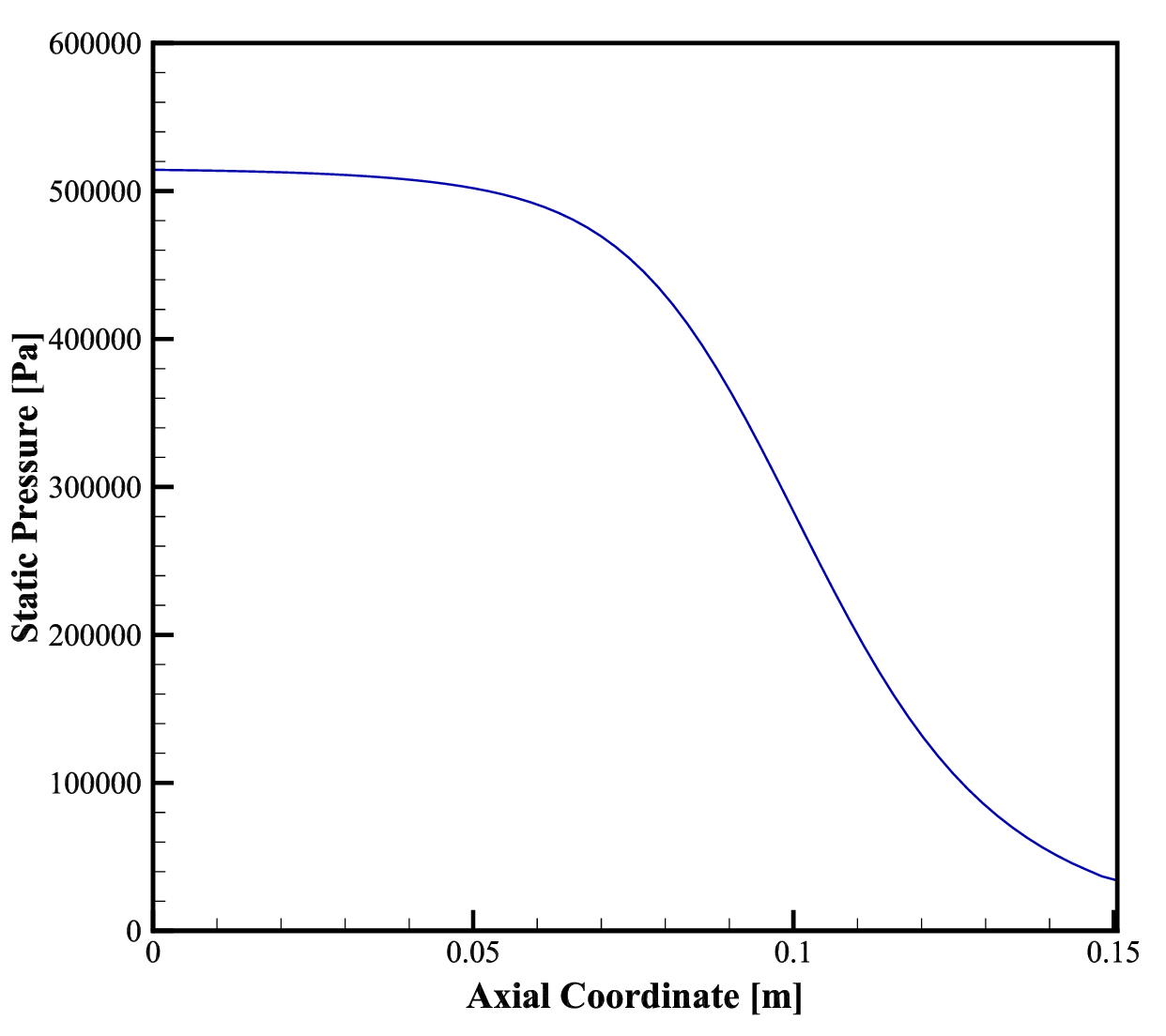}
        \label{fig:Pressure variation along the axis}
    }
    \caption{Pressure variation.}
\end{figure}

The flow properties like mach number Fig. \ref{fig:Mach Contour.}, \ref{fig:Mach Number Along the axis}, temperature Fig. \ref{fig:Temperature Contour}, \ref{fig:Temperature variation along the axis} and pressure Fig. \ref{fig:Pressure Contour}, \ref{fig:Pressure variation along the axis} with their respective axial variation can be visualized in the plots provided.  


\section{Results And Discussion}

\begin{figure}[hbt!]
\centering
\includegraphics[width=0.6\textwidth]{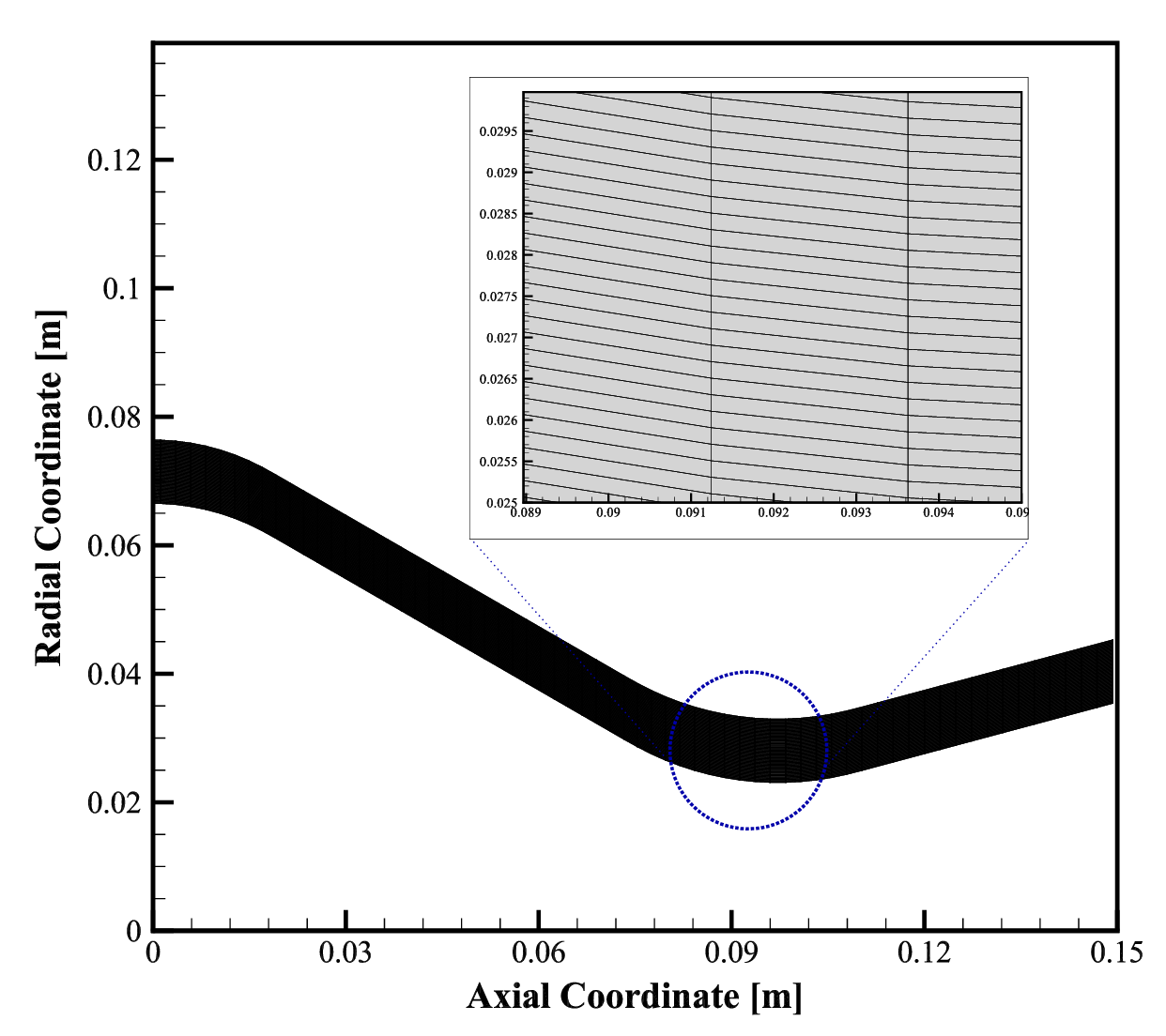}
\caption{Mesh for the thermal protection system}
\label{fig:TPSMesh} 
\end{figure}

After successful validation of flow solver, complete conjugate heat transfer analysis is performed. This time a thermal protection system (TPS) is introduced on the nozzle wall which is analyzed for the solid domain. TPS is made of AVCOAT 5026–39/HC-G, a glass-filled epoxy novolac ablative material. The thickness of 10 mm is provided for the solid domain simulations. AVCOAT material has an ablation temperature of 922 K and the relative thermo-physical properties of the material are taken from Ref.~\cite{williams1992thermal}. The mesh used for the flow simulations is same as used for validation case study, shown in Sec.~\ref{Validation}. The mesh of the solid domain, as shown in Fig. \ref{fig:TPSMesh}, is having 250 cells in transverse direction and 50 cells in radial direction. The boundary conditions for the further study are summarized in Table~\ref{tab:boundary_conditions}. Initially, an isothermal boundary condition was imposed on the nozzle wall, with the wall temperature set to half the nozzle inlet temperature at the initial time step. As the simulation progressed, a user-defined function (UDF) was employed to apply a spatially varying temperature distribution on the nozzle wall.

\begin{table}[hbt!]
    \centering
        {\caption{Boundary conditions and operating parameters for the nozzle with Thermal Protection System (TPS).}%
        \label{tab:boundary_conditions}}%
        {%
        \begin{tabular}{|l|c|}
        \hline
        \textbf{Condition}                    & \textbf{Value}       \\ \hline
        Nozzle inlet pressure                 & 344048.40 N/m\(^2\) \\ \hline
        Nozzle inlet temperature              & 1111.11 K             \\ \hline
        Nozzle inlet density                  & 1.07 kg/m\(^3\)    \\ \hline
        Transport model                       & Sutherland’s law     \\ \hline
        Turbulence model                      & BSL                  \\ \hline
        Exit Mach number                      & 2.5                  \\ \hline
        \end{tabular}
        }
\end{table}

Simulation results were evaluated at various time intervals to capture the thermal response of the nozzle with the TPS layer, providing insights into the material's performance under operational conditions.

\subsection{Comparison of Heat Flux and Surface Temperature}

\begin{figure}[hbt!]
    \centering
    \subfloat[Heat flux variation from 0.2–1.0 s]{
        \includegraphics[width=0.48\textwidth]{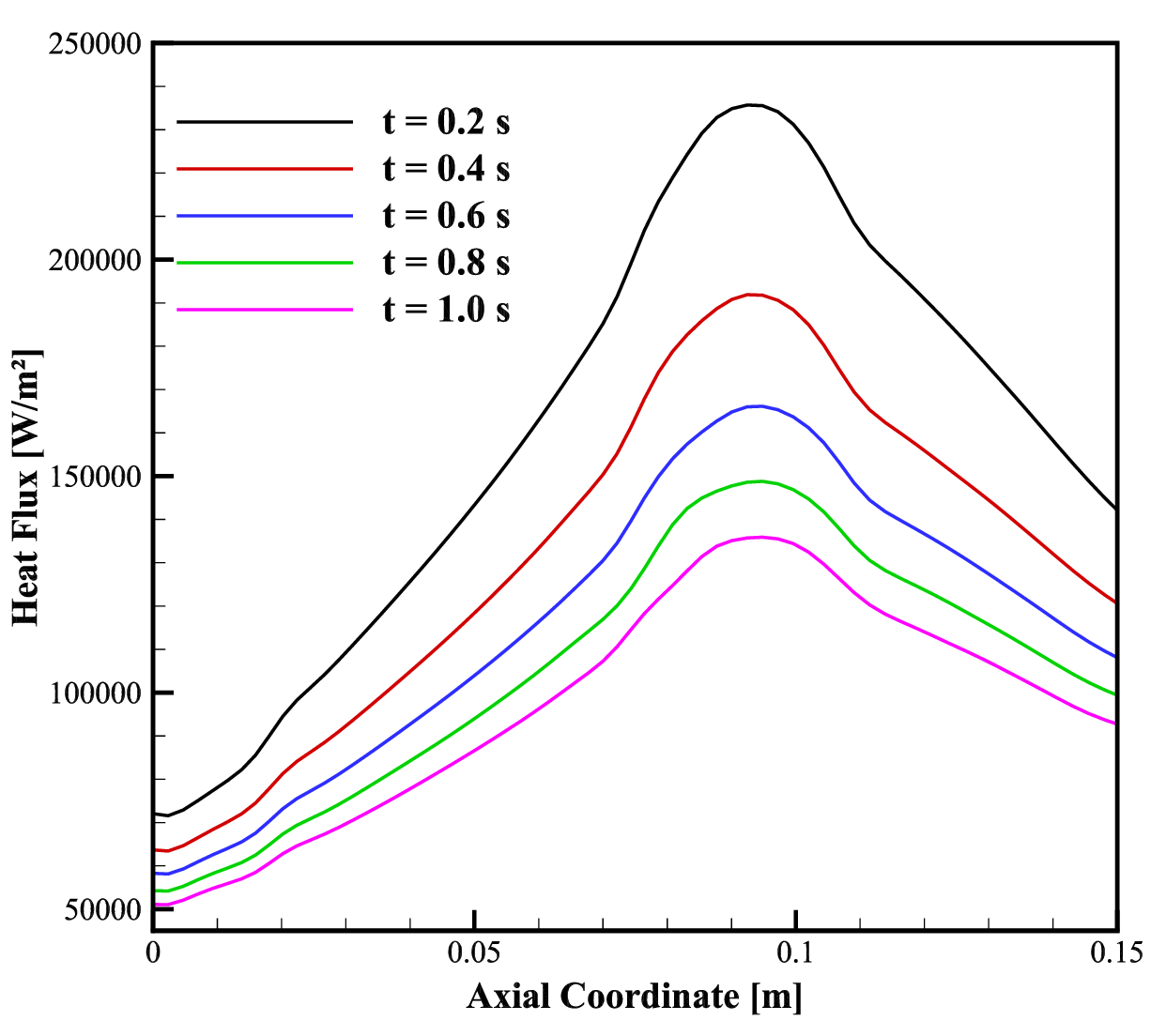}
        \label{fig:Heat Flux Variation0.2-1.}
    }
    \hfill
    \subfloat[Surface temperature at 1.0 s]{
        \includegraphics[width=0.48\textwidth]{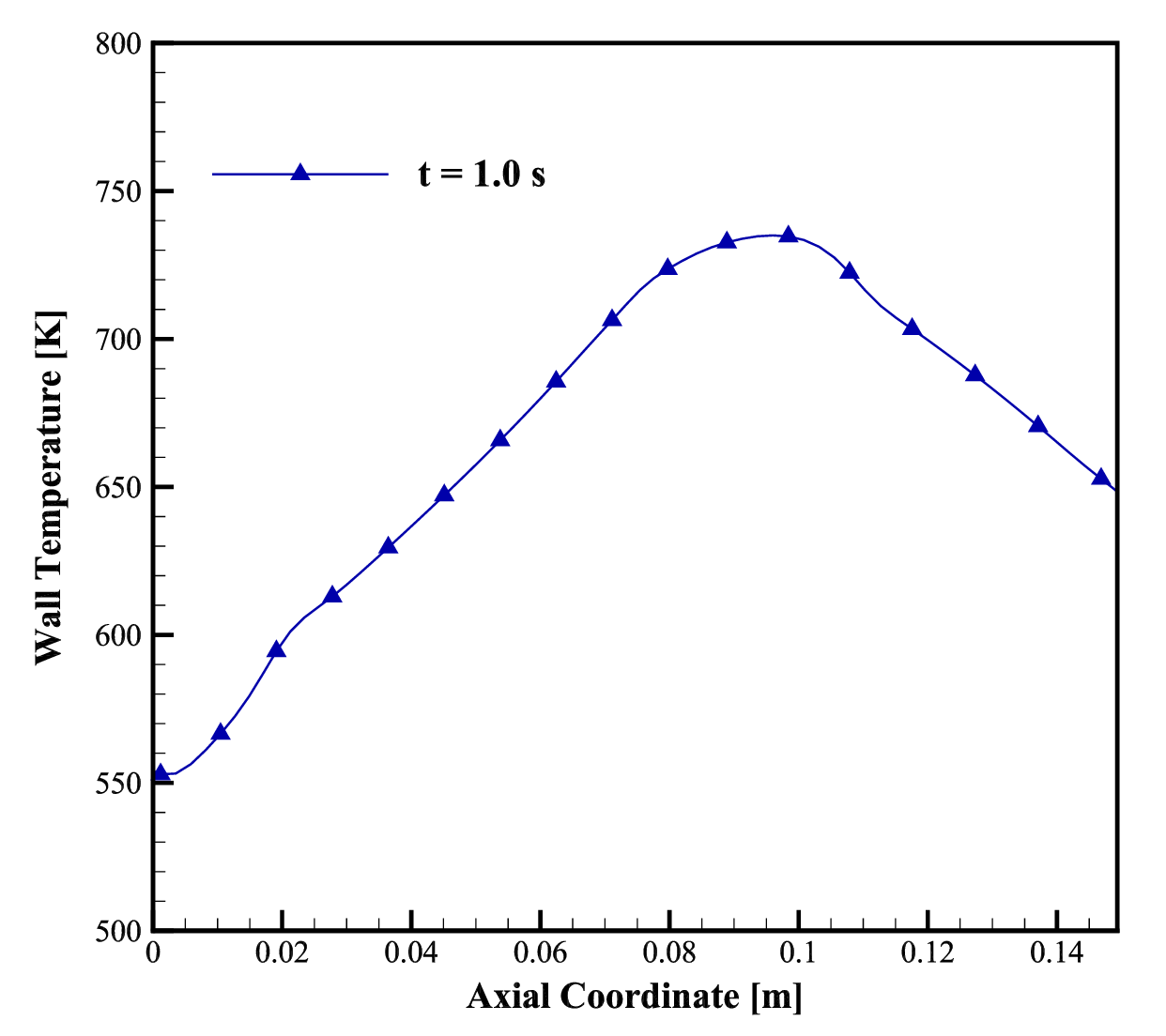}
        \label{fig:TPS_TEMP_1_sec}
    }
    \caption{Comparison of surface heat flux (0.2–1.0 s) and surface temperature (1.0 s).}
\end{figure}

Figure~\ref{fig:Heat Flux Variation0.2-1.} shows the convective heat flux variation along the nozzle during the first second of operation. The same heat flux is applied to the TPS material to evaluate the thermal response of the AVCOAT material. It can be observed that the maximum heat flux occurs in the throat region of the nozzle. This corresponds to the maximum surface temperature in this region, as can be seen from Fig. \ref{fig:TPS_TEMP_1_sec}. Additionally, it can be observed that the heat flux shows a decreasing trend with increasing time in this time interval.

\begin{figure}[hbt!]
    \centering
    \subfloat[Heat flux variation from 1.0-5.0 s]{
        \includegraphics[width=0.48\textwidth]{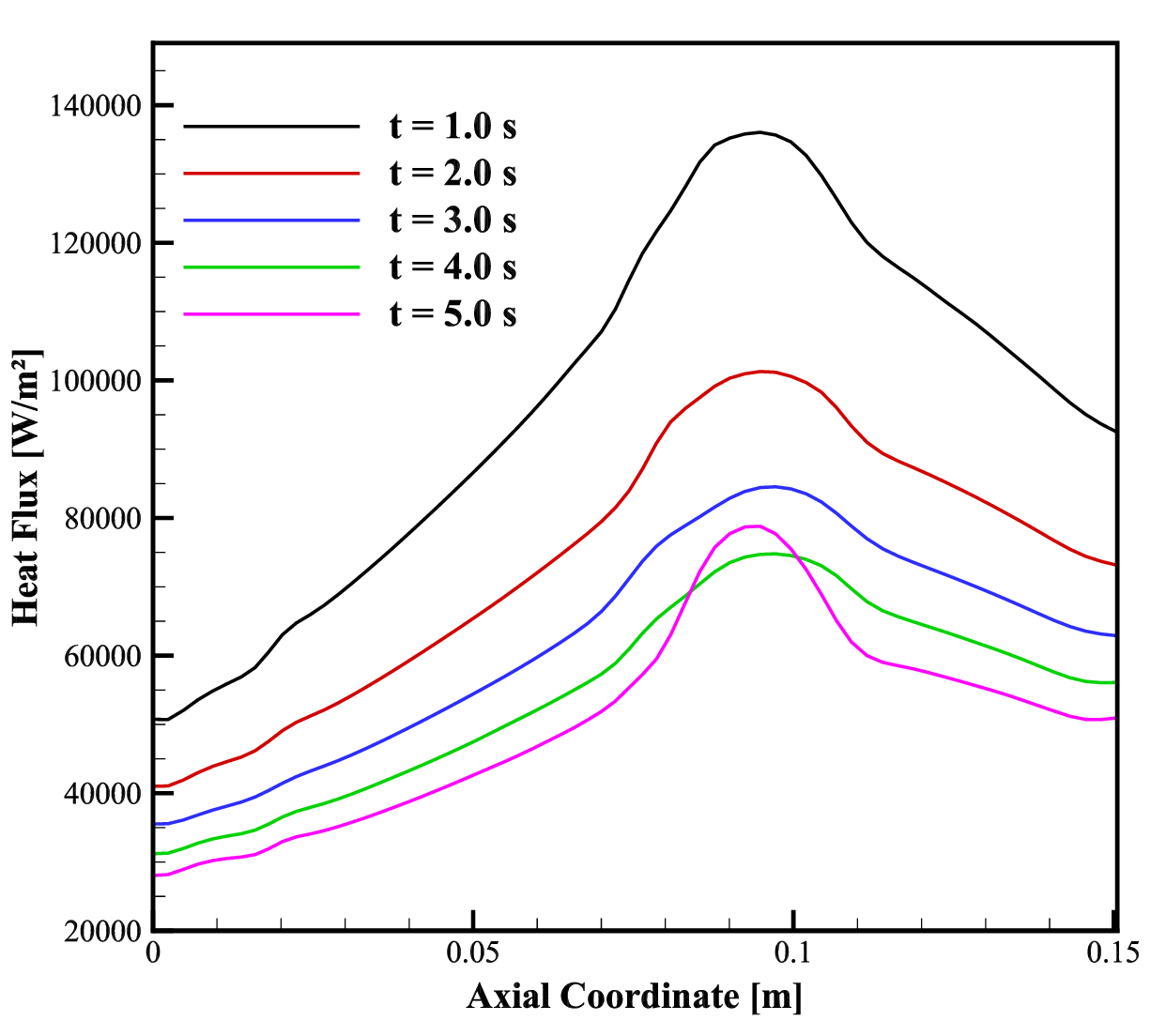}
        \label{fig:Heat Flux Variation1-5.}
    }
    \hfill
    \subfloat[Surface temperature variation at 5.0 s]{
        \includegraphics[width=0.48\textwidth]{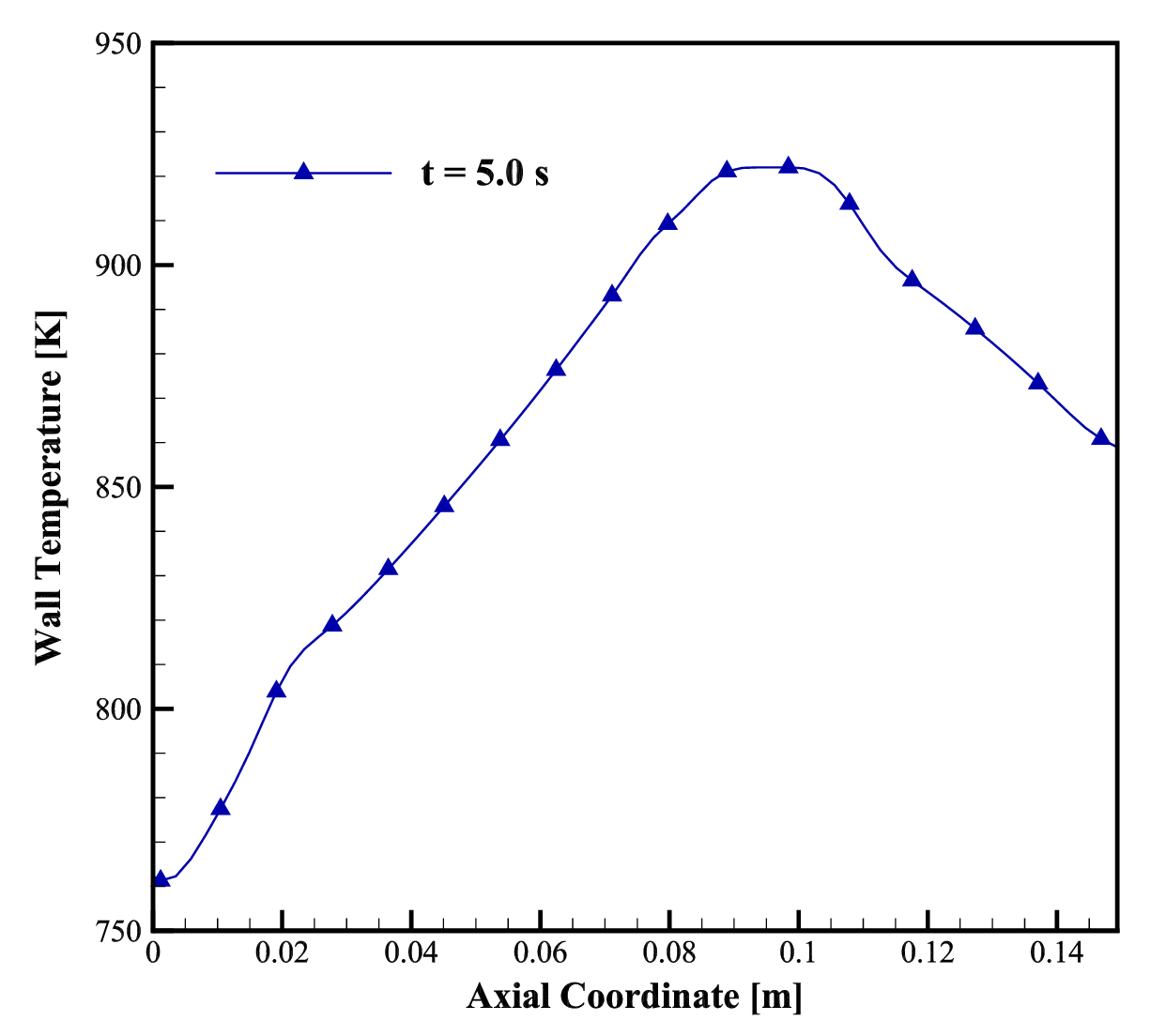}
        \label{fig:TPS_TEMP_1-5}
    }
    \caption{Comparison of surface heat flux (1.0-5.0 s) and surface temperature (5.0 s).}
\end{figure}

Figure~\ref{fig:Heat Flux Variation1-5.} shows the temporal evolution of the wall heat flux at the nozzle throat over the interval 1--5\,s. The heat flux decreases during the early portion of this interval but reverses and increases during the fifth second. This behavior is examined in detail later in Sec.~\ref{subsection:thermal_field_characteristics} and is attributed to an increase in near-wall viscous dissipation together with changes in the TPS surface temperature. As shown in Fig.~\ref{fig:TPS_TEMP_1-5}, the TPS surface temperature at the throat attains the ablation threshold (922\,K) during the fifth second, indicating that ablation initiates earlier at the throat than at other axial locations. The onset of ablation and the resulting modification of the surface boundary condition steepen the local wall-normal temperature gradient, which explains the observed rise in instantaneous wall heat flux.

\begin{figure}[hbt!]
    \centering
    \subfloat[Heat flux variation from 5.0-10.0 s]{
        \includegraphics[width=0.48\textwidth]{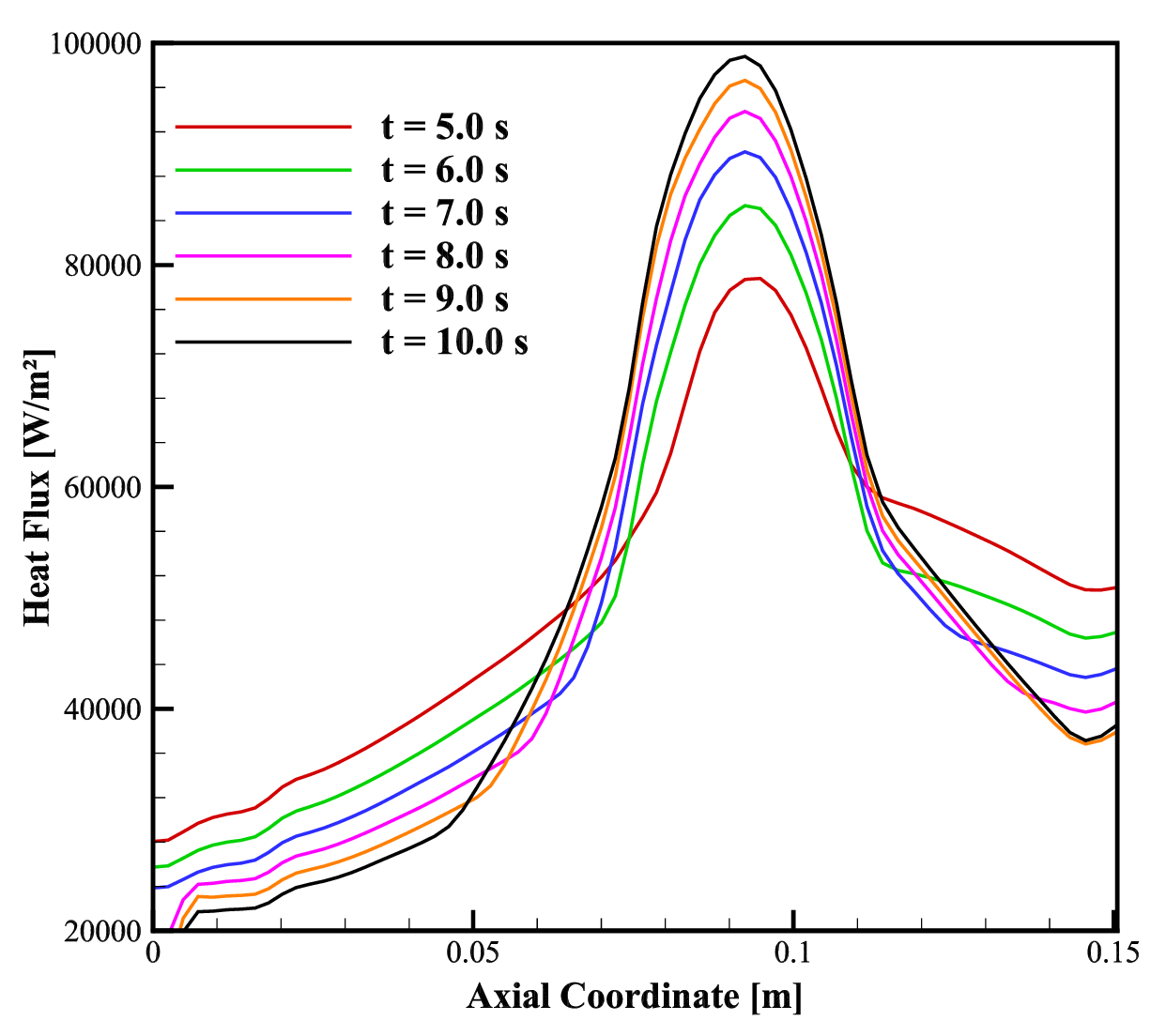}
        \label{fig:Heat Flux variation5-10}
    }
    \hfill
    \subfloat[Surface temperature variation at 10.0 s]{
        \includegraphics[width=0.48\textwidth]{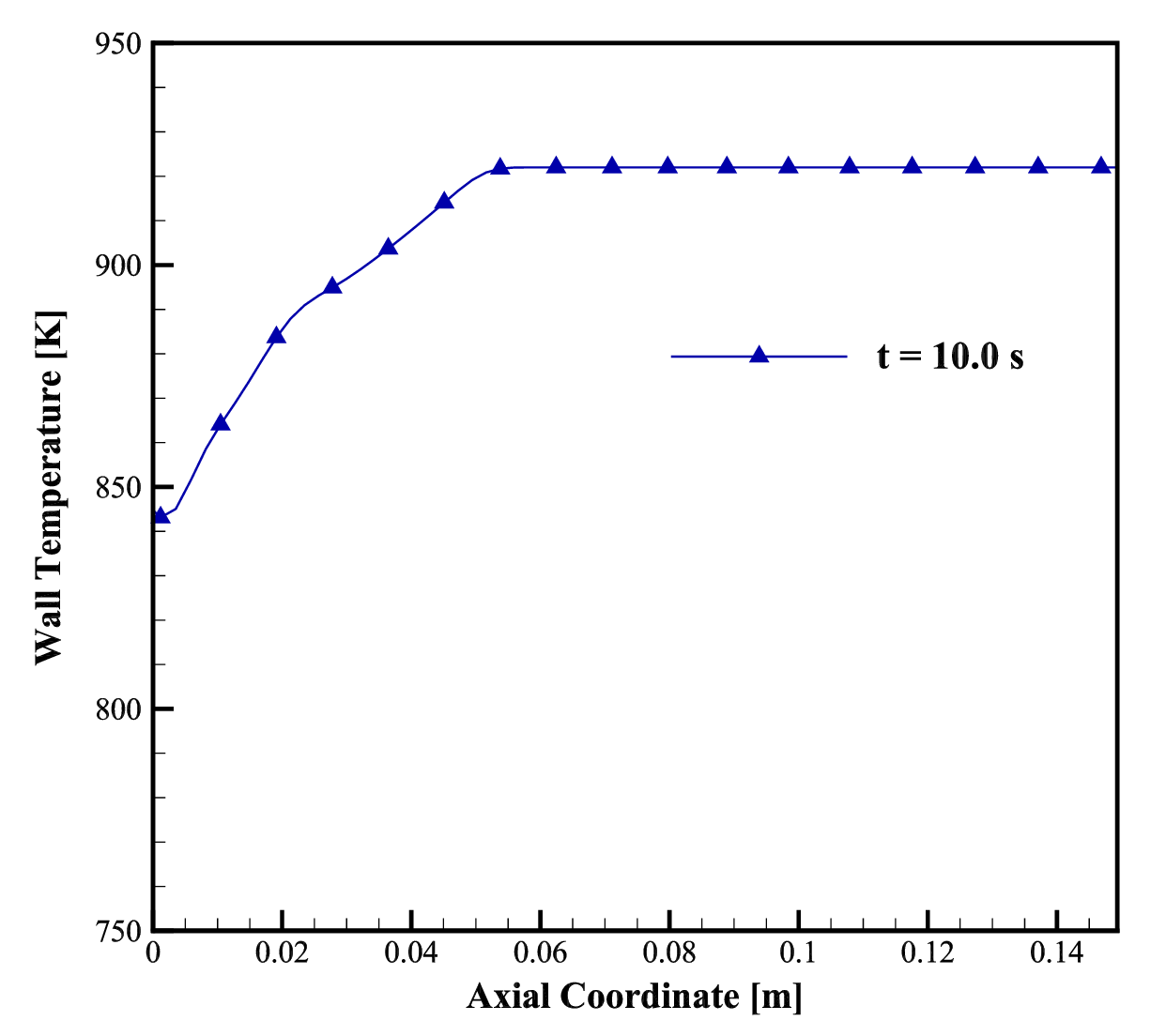}
        \label{fig:TPS_TEMP_5-10sec}
    }
    \caption{Comparison of heat flux (5.0-10.0 s) and surface temperature (10.0 s).}
\end{figure}

During the time interval of 5 to 10 s, the heat flux showed a complete reversal in heat flux trend, as depicted in Fig.~\ref{fig:Heat Flux variation5-10}. A gradual and consistent increment in the heat flux is evident over time. From the temperature profile, Fig.~\ref{fig:TPS_TEMP_5-10sec}, it can be seen that a large portion of the nozzle wall has reached the ablation threshold of the material. Furthermore, the results illustrate that the ablation is spreading to the adjacent regions of the throat area as specified by the expansion of both heat flux and temperature profiles.

\begin{figure}[hbt!]
    \centering
    \subfloat[Heat flux variation from 11.0-15.0 s]{
        \includegraphics[width=0.48\textwidth]{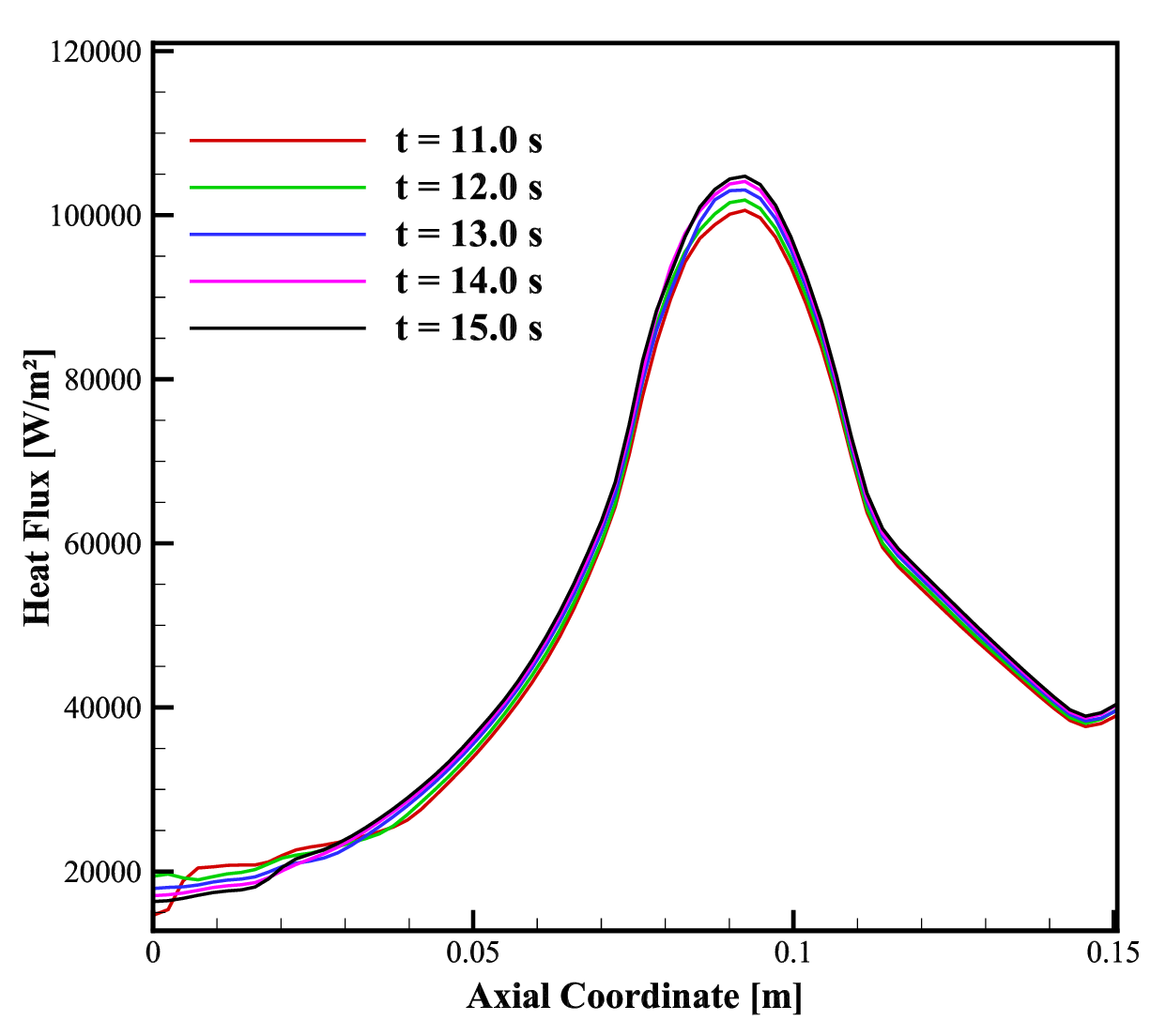}
        \label{fig:Heat Flux variation10-15}
    }
    \hfill
    \subfloat[Surface temperature variation at 15.0 s]{
        \includegraphics[width=0.48\textwidth]{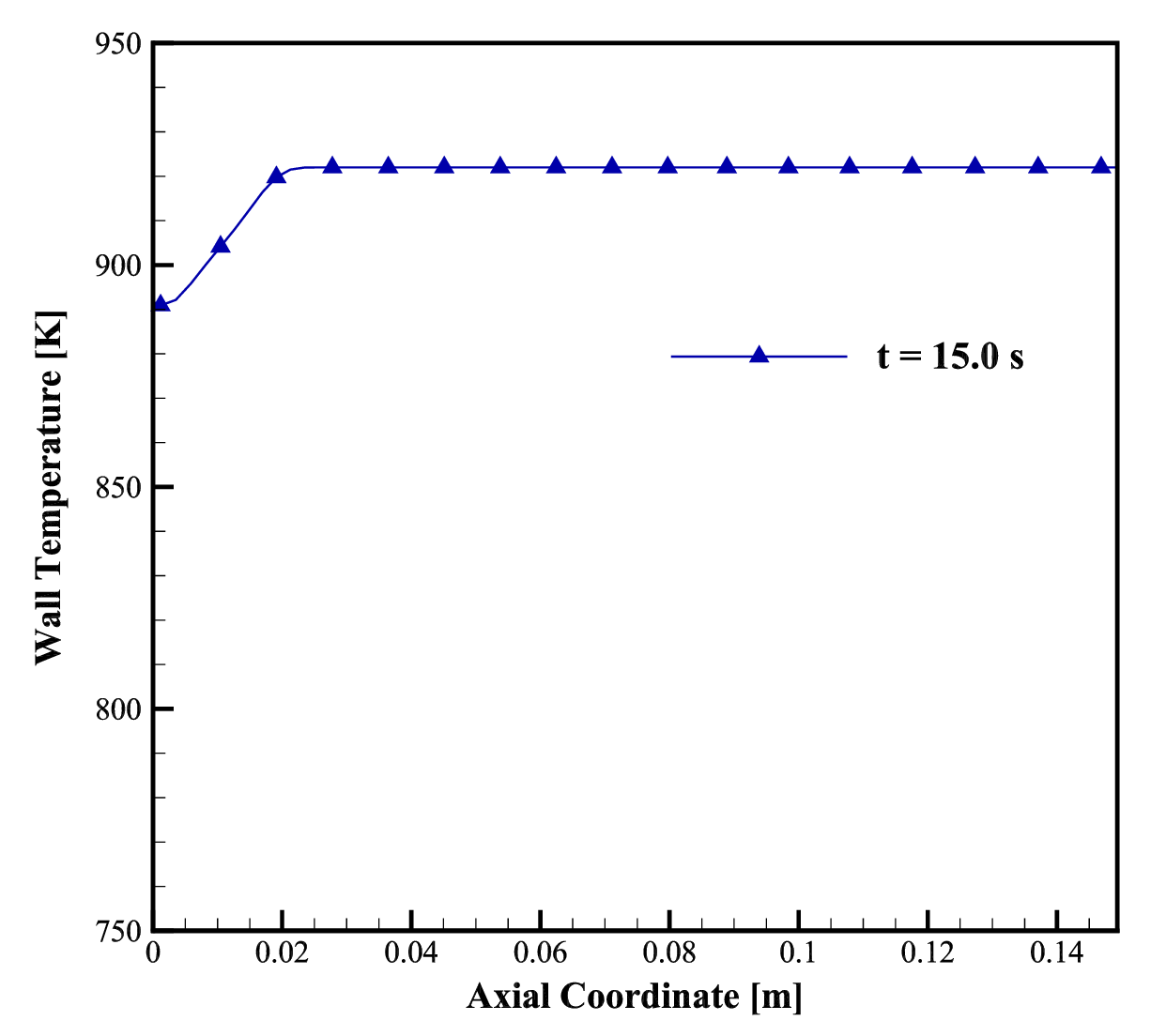}
        \label{fig:TPS_TEMP_10-15sec}
    }
    \caption{Comparison of heat flux (11.0-15.0 s) and surface temperature (15 s).}
\end{figure}

Between 11 and 15 s, the heat flux stabilizes across the nozzle wall. It becomes apparent that the increment in the heat flux is gradual, indicating it is approaching a steady state, as shown in Fig. \ref{fig:Heat Flux variation10-15}. From the temperature profile (Fig.~\ref{fig:TPS_TEMP_10-15sec}), it is clear that the most part of nozzle wall has undergone ablation during this time interval. 

\begin{figure}[hbt!]
    \centering
    \subfloat[Heat flux variation from 16.0-20.0 s]{
        \includegraphics[width=0.48\textwidth]{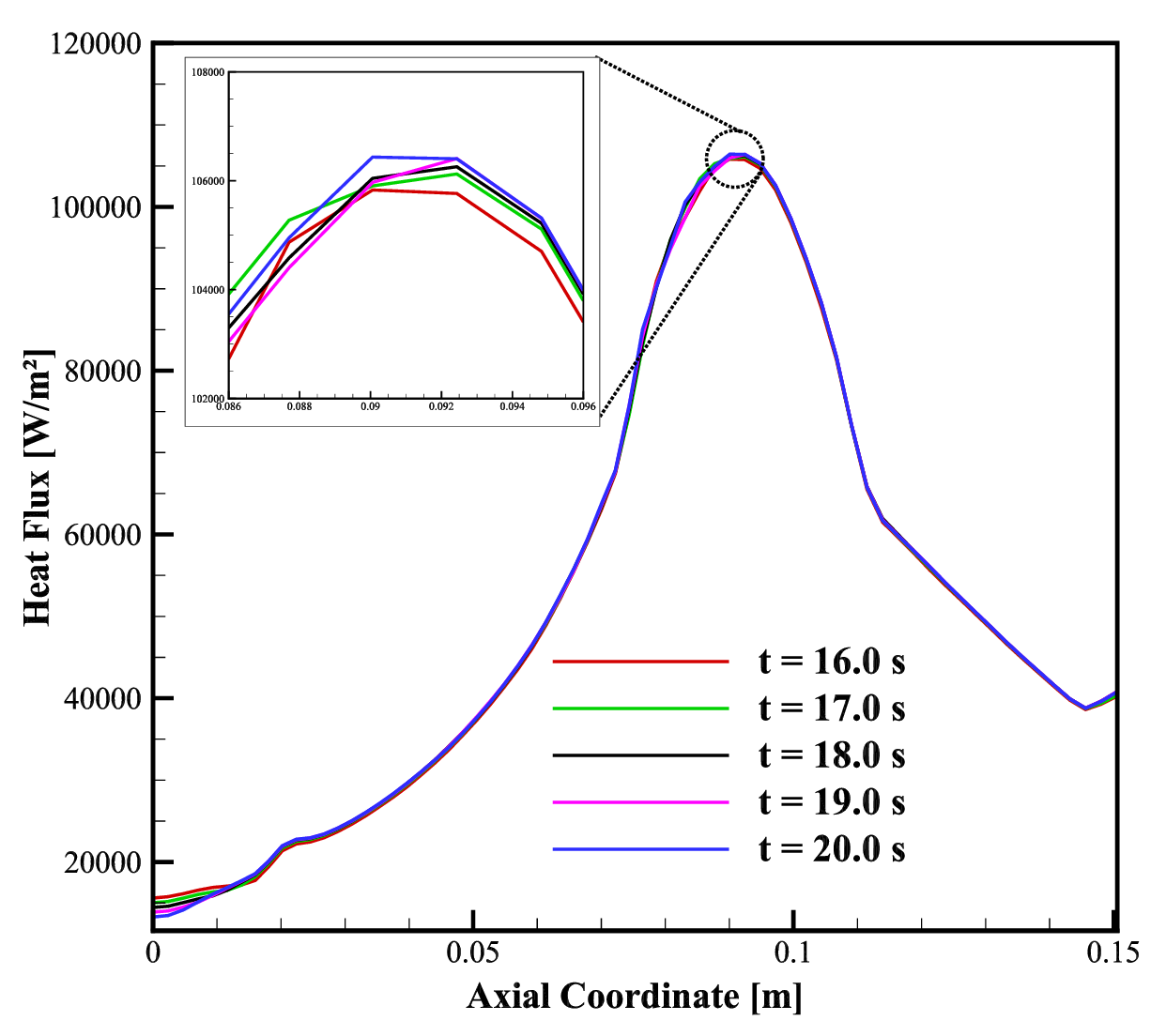}
        \label{fig:Heat Flux variation15-20.}
    }
    \hfill
    \subfloat[Surface temperature variation at 20.0 s]{
        \includegraphics[width=0.48\textwidth]{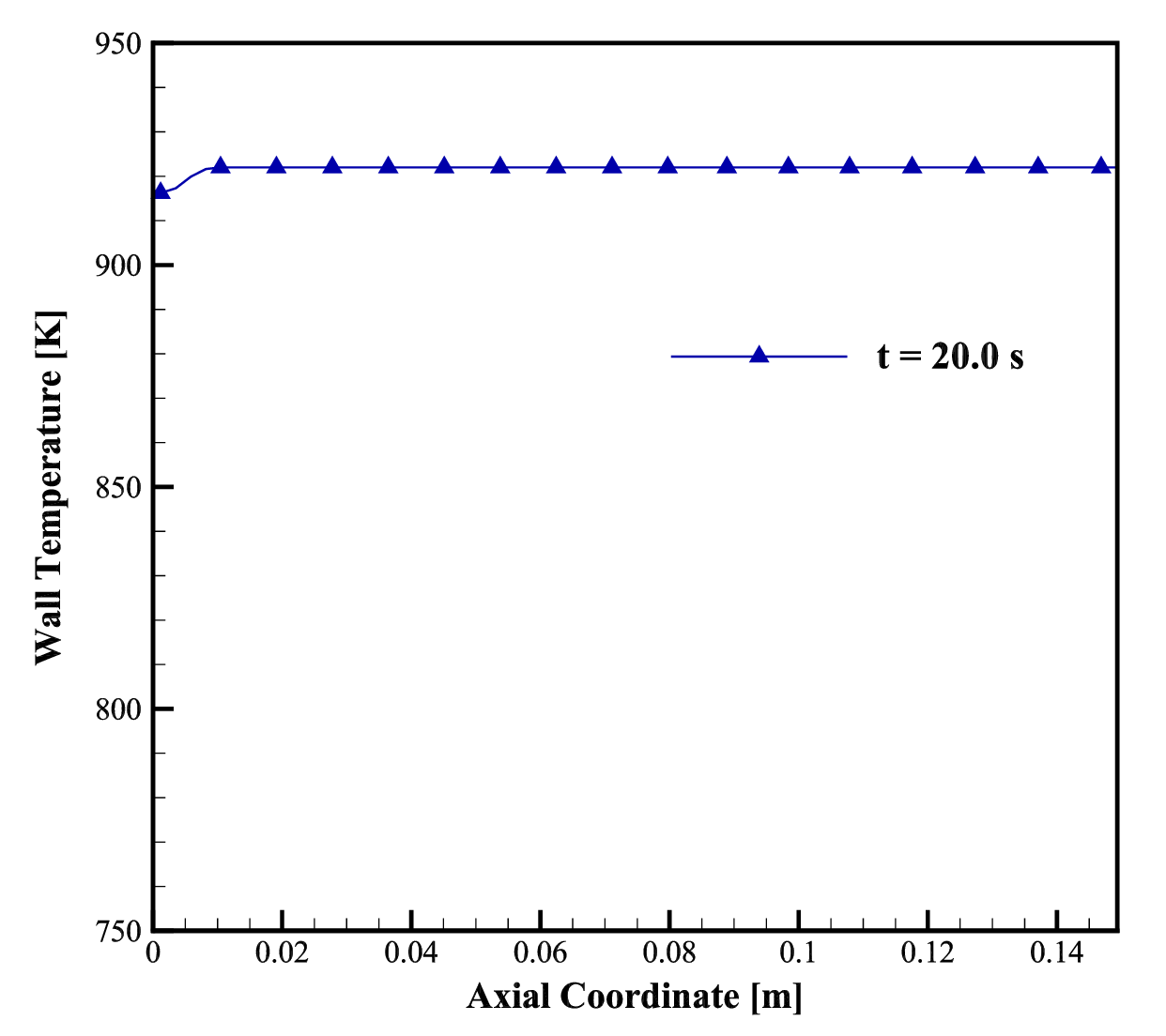}
        \label{fig:TPS_TEMP_15-20sec}
    }
    \caption{Comparison of heat flux (16.0-20.0 s) and surface temperature (20.0 s).}
\end{figure}

During the 16-20 s interval, the progression in the heat flux is minimal, as illustrated in Fig.~\ref{fig:Heat Flux variation15-20.}. The surface temperature of the nozzle wall, with the exception of entry section reached the ablation threshold of 922 K, except the entry part of the nozzle, as highlighted in Fig.~\ref{fig:TPS_TEMP_15-20sec}

\begin{figure}[hbt!]
    \centering
    \subfloat[Heat flux variation from 20.0-25.0 s]{
        \includegraphics[width=0.48\textwidth]{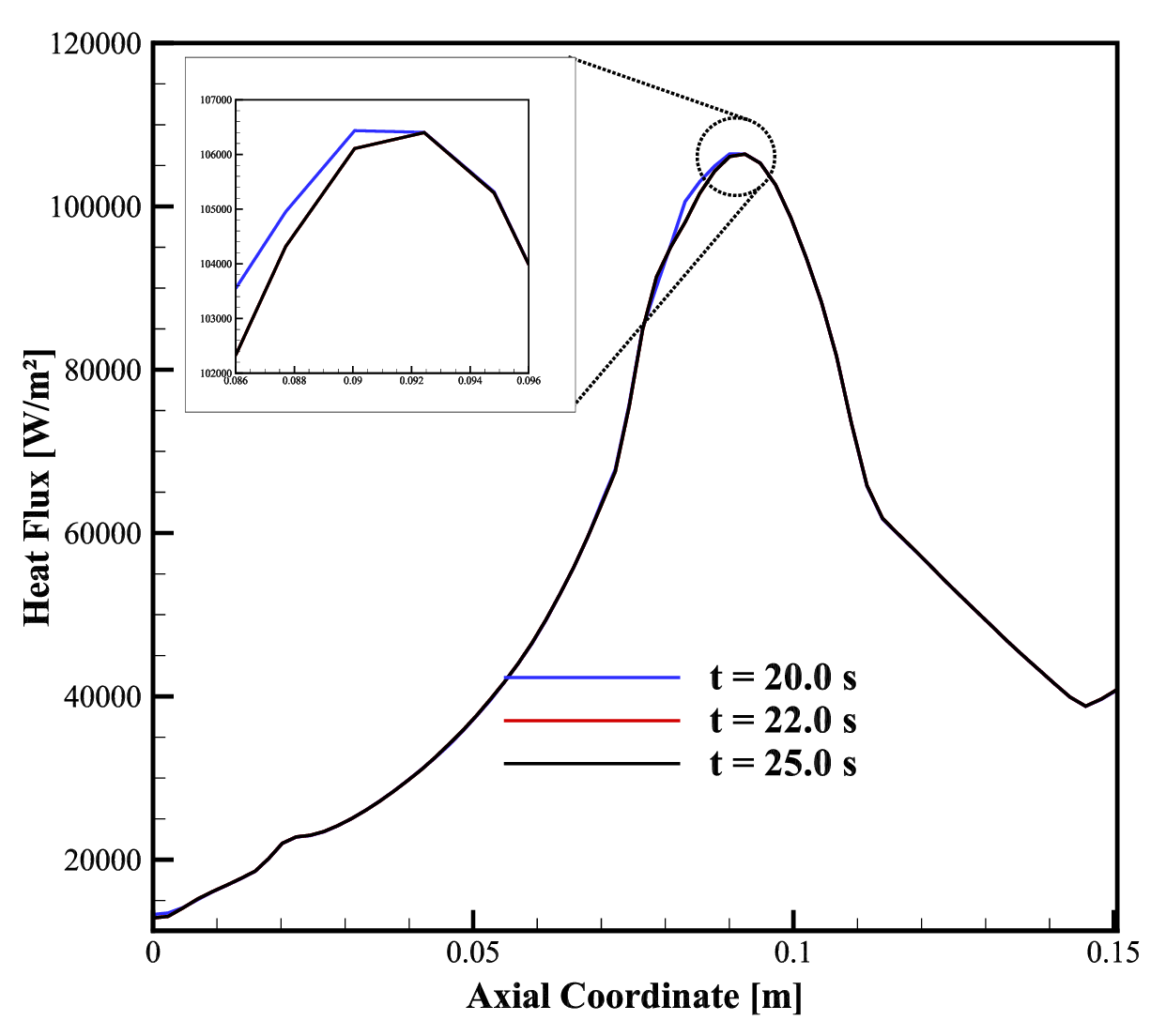}
        \label{fig:Heat Flux variation20-25}
    }
    \hfill
    \subfloat[Surface temperature variation at 25.0 s]{
        \includegraphics[width=0.48\textwidth]{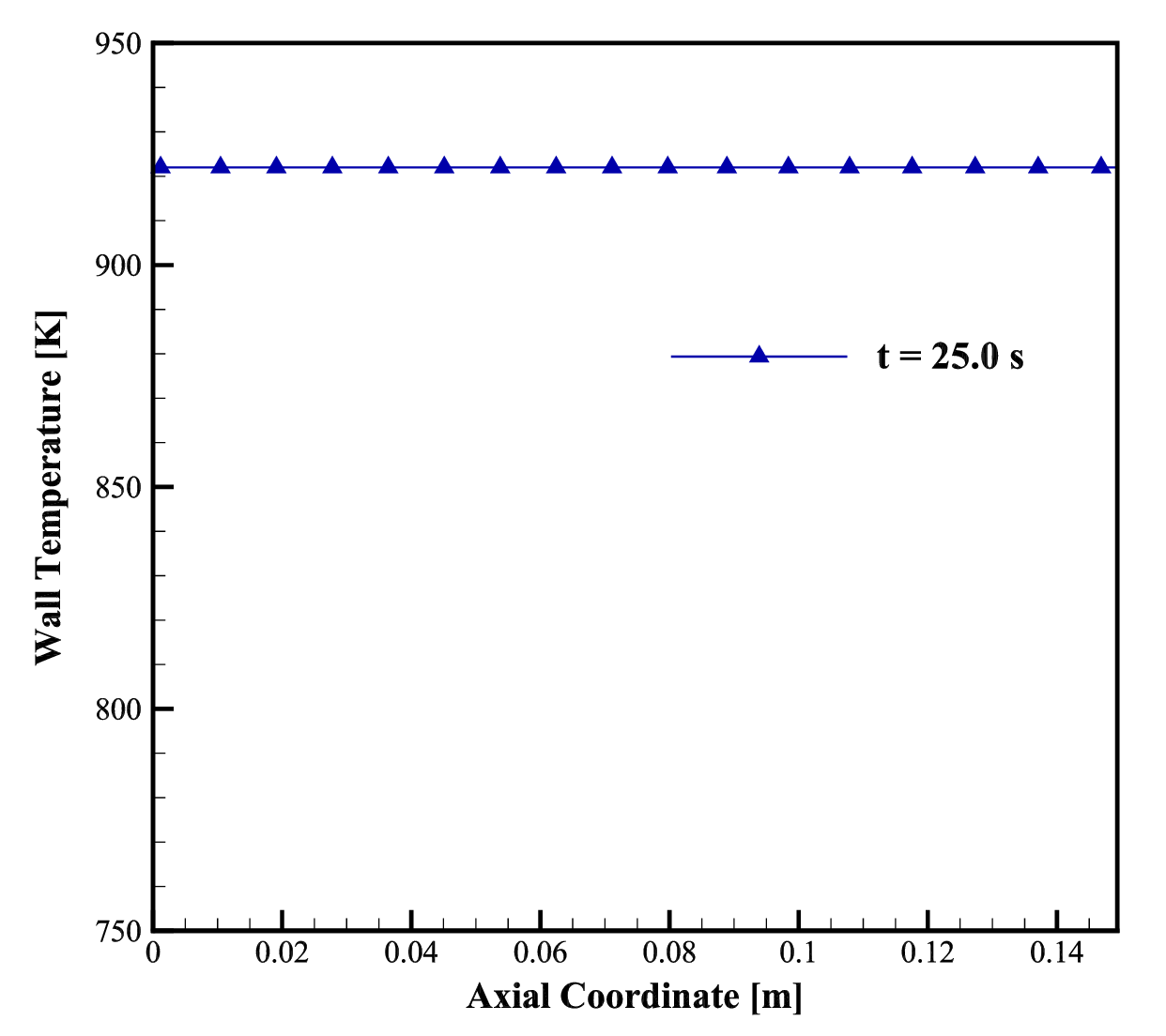}
        \label{fig:TPS_TEMP_20-25sec}
    }
    \caption{Comparison of heat flux (20.0-25.0 s) and surface temperature (25 s)}
\end{figure}

As presented in Fig.~\ref{fig:Heat Flux variation20-25}, the heat flux in the interval of 20 to 25 s reaches a stable state. There is no further increase in the heat flux over time. From the Fig.~\ref{fig:Heat Flux variation20-25}, it is clear that the heat flux at 22 s and 25 s is identical. The surface temperature of the nozzle wall reaches an isothermal temperature 922 K in this time, as observed in Fig. \ref{fig:TPS_TEMP_20-25sec}. 

Simulation results were extended further in time, up to 120 s. Since the nozzle wall temperature remained at 922 K, no further increment in the heat flux was evident. The heat flux at 120 s matched completely with that at 22 s. At this point, the entire nozzle wall has undergone ablation.

\subsection{Throat Region Heat Transfer Analysis}
\label{subsection:thermal_field_characteristics}
\begin{figure}[hbt!]
    \centering
    \subfloat[Nozzle throat heat flux variation]{
        \includegraphics[width=0.48\textwidth]{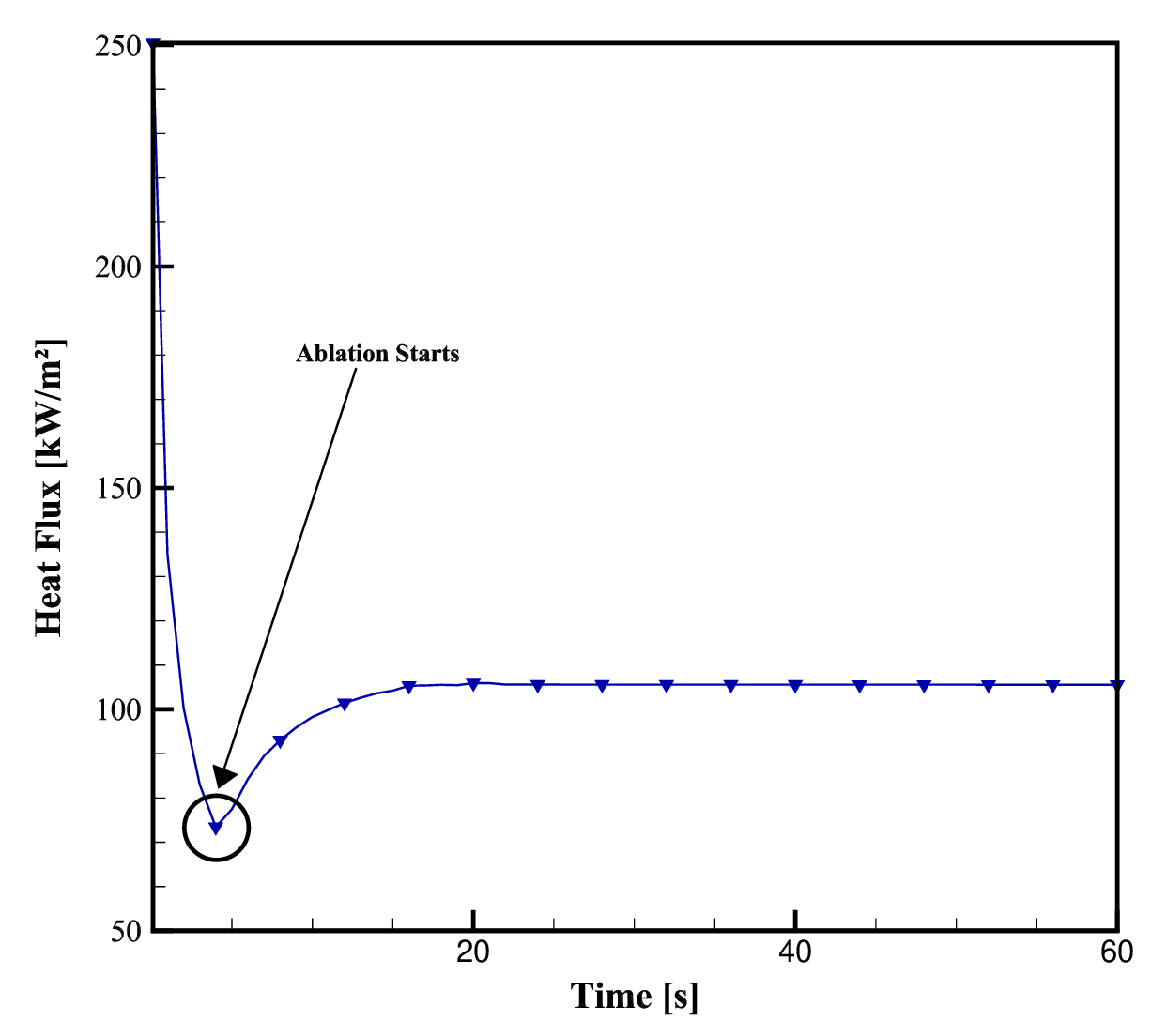}
        \label{fig:Nozzle throat Heat Flux variation}
    }
    \hfill
    \subfloat[Nozzle throat temperature variation]{
        \includegraphics[width=0.48\textwidth]{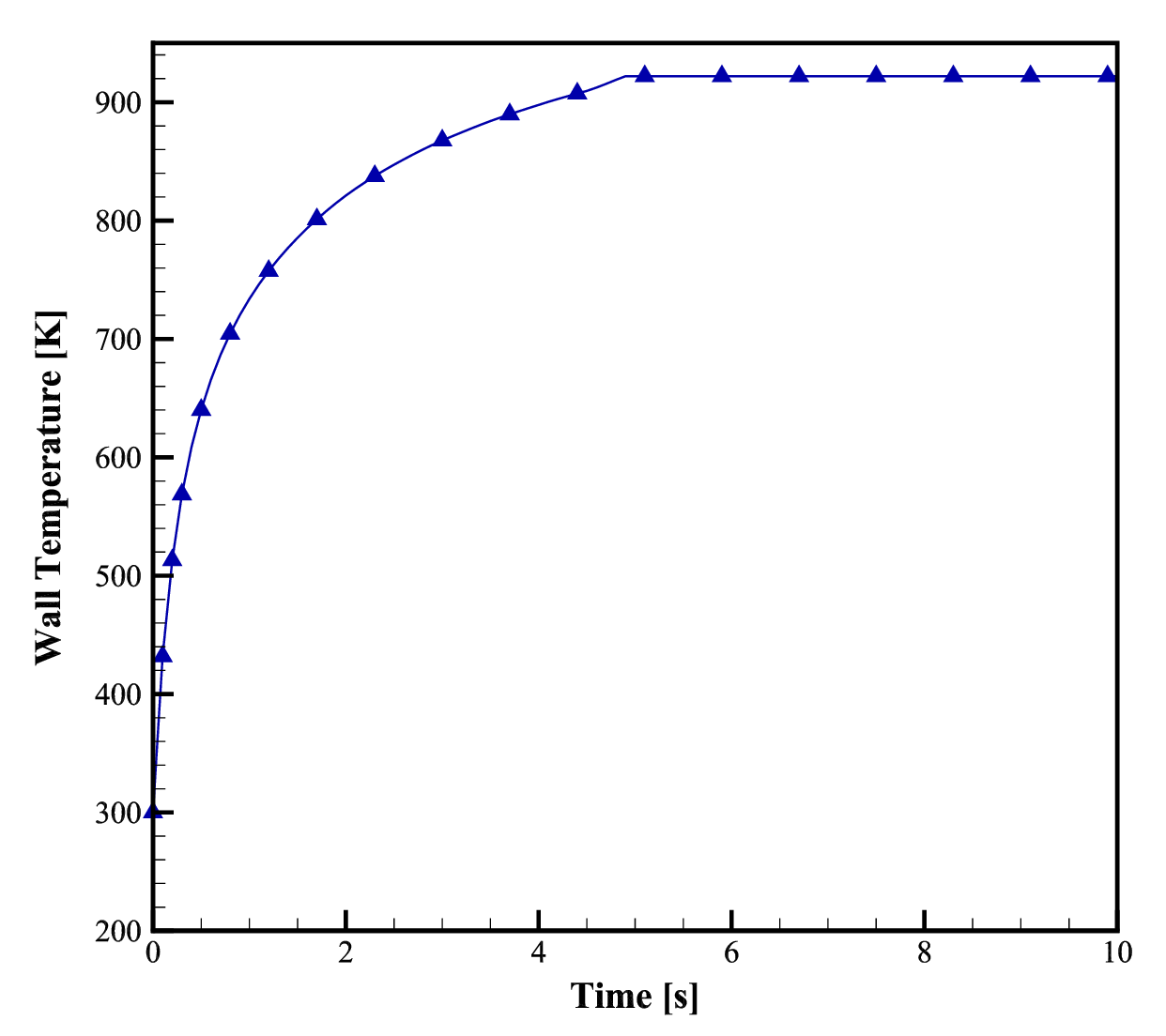}
        \label{fig:Nozzle Throat temperature variation}
    }
    \caption{Comparison of heat flux and temperature at nozzle throat}
\end{figure}

The throat of a rocket nozzle is a critical region as it separates the subsonic flow region from the supersonic flow, hence a detailed analysis is needed. This region of the nozzle experiences maximum heat flux, as discussed previously. The results of heat flux variation with time for the throat region are highlighted in Fig.~\ref{fig:Nozzle throat Heat Flux variation}. The results illustrate that initially the heat flux is at its highest value, followed by a sudden decrease in the heat flux as time progresses. As demonstrated in the graph, the ablation process begins at around 5 s, at which point the heat flux increases again for a few seconds. Further in time, one can see that the heat flux attains a stable value.

In comparison to the heat flux, the surface temperature at the throat continues to increase with time until it reaches the ablation temperature of 922\,K, as shown in Fig.~\ref{fig:Nozzle Throat temperature variation}. Once the throat surface attains the ablation temperature, the temperature remains fixed thereafter. Simulations were carried out up to 120\,s, but results are displayed only up to 60\,s for heat flux and 10\,s for surface temperature, since no further changes in these quantities were observed beyond those times.

\subsection{Thermal Field Characteristics}

To interpret the observed heat flux trends, it is first necessary to examine the local flow properties near the nozzle throat. For this purpose, the static temperature was evaluated at the throat to analyze its radial variation from the nozzle centerline ($y=0$) to the cooled wall.

\begin{figure}[hbt!]
    \centering
    \subfloat[Radial distribution of static temperature at the nozzle throat.]{
        \includegraphics[width=0.48\textwidth]{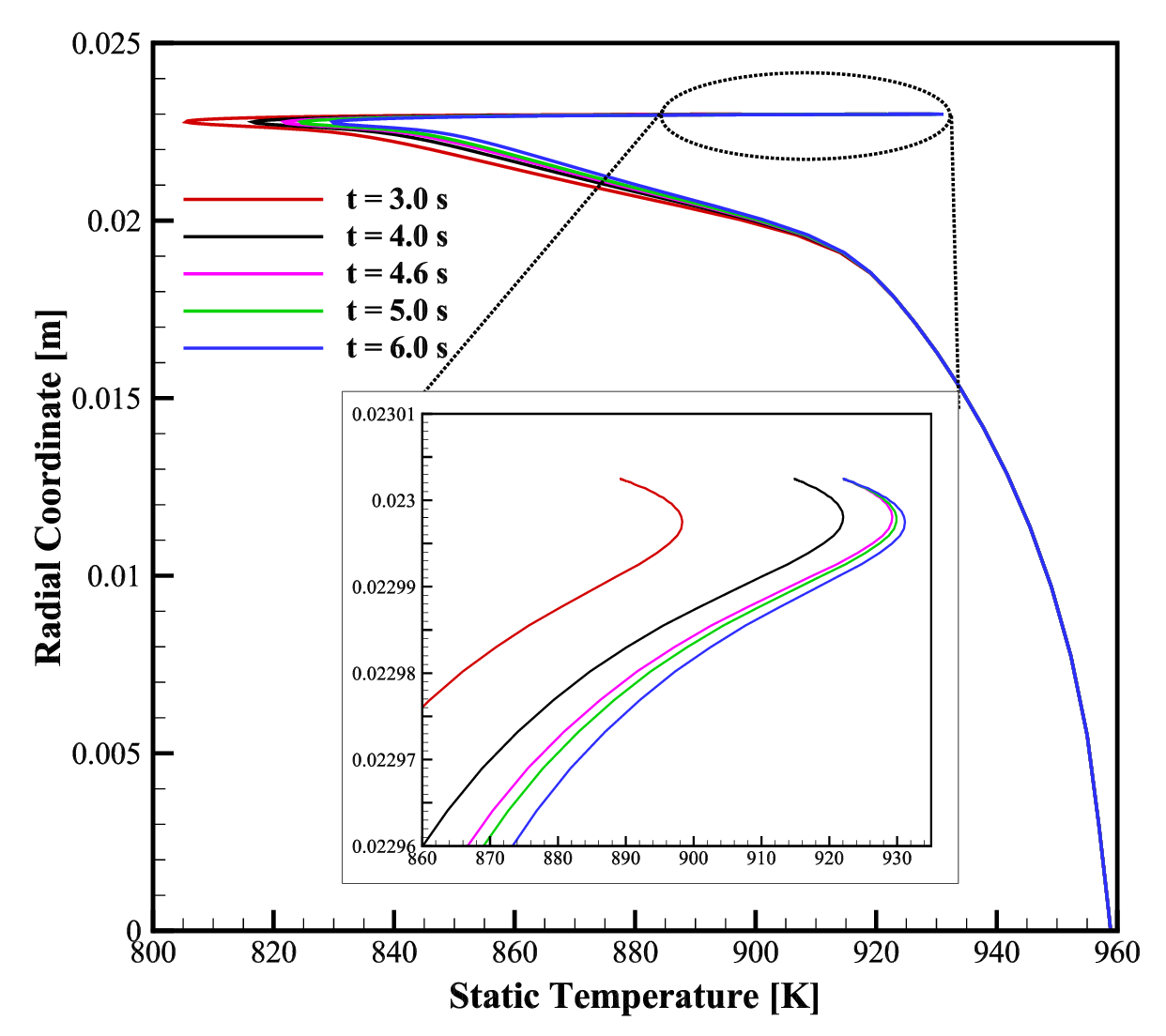}
        \label{fig:throat_static_temperature}
    }
    \hfill
    \subfloat[Temperature gradients at the nozzle throat.]{
        \includegraphics[width=0.48\textwidth]{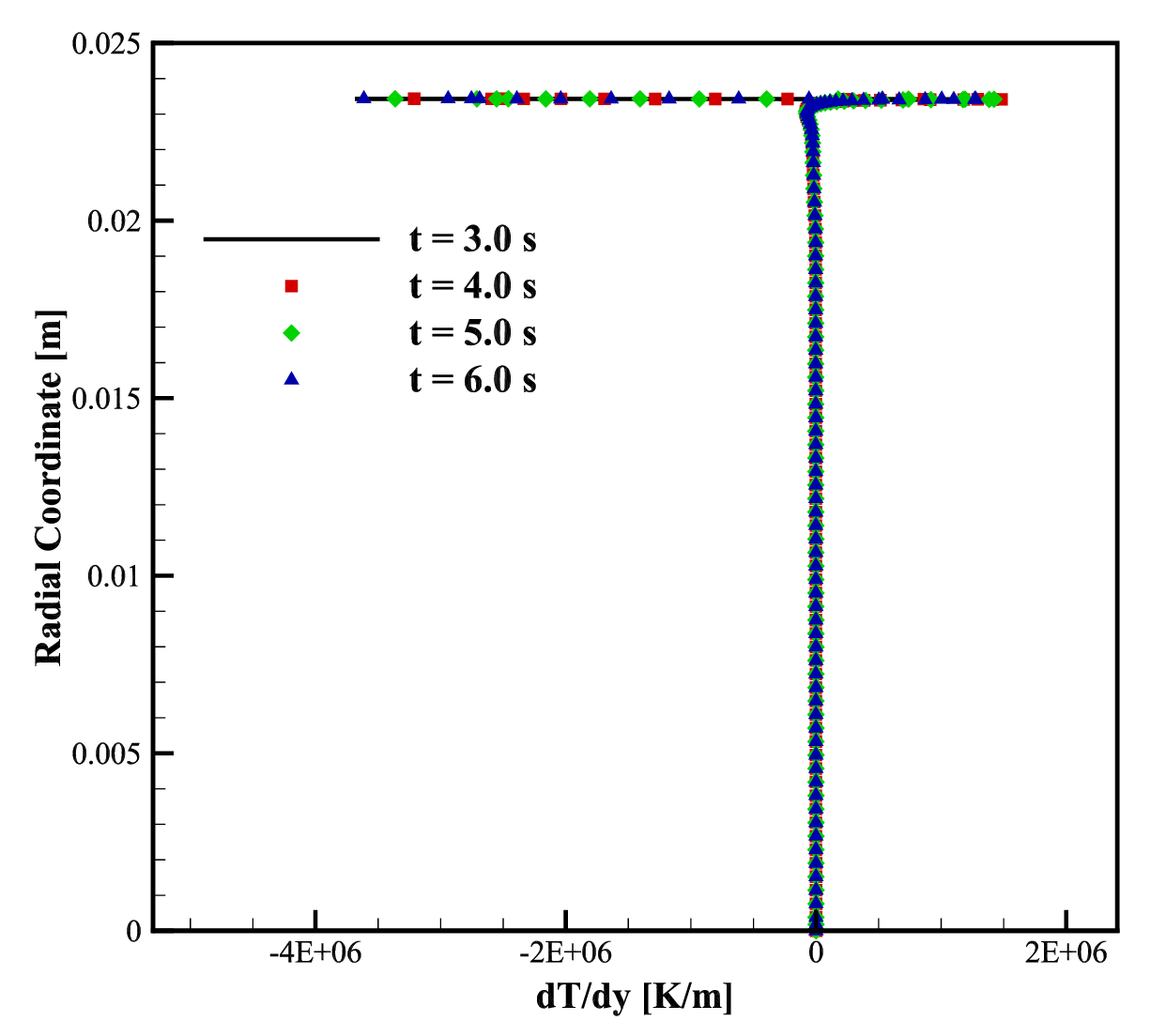}
        \label{fig:Temperature gradients}
    }
    \caption{Static temperature and temperature gradients across the nozzle throat at different times.}
    \label{fig:throat_temperatures}
\end{figure}

Figure~\ref{fig:throat_static_temperature} shows the radial static temperature distribution over the interval 3--6\,s. This time interval was chosen to capture the flow properties before and after ablation initiates. The temperature profile exhibits three distinct regions: an initial gradual decrease, a local rise, and a final drop to the wall temperature.

In the outer portion of the boundary layer, mixing with slower and slightly cooler fluid, together with heat transfer toward the cooled wall, produces a modest reduction in temperature relative to the core ($y=0$). Closer to the wall, within the inner boundary layer, the steep reduction in axial velocity generates strong velocity gradients. The associated viscous dissipation converts kinetic energy into internal energy, producing a local recovery effect that exceeds the immediate conductive losses. This mechanism accounts for the observed rise in the temperature profile. The gas temperature is constrained by the prescribed cold-wall boundary condition; since the wall temperature is maintained below the recovery temperature, the static temperature falls sharply, reaching the prescribed surface temperature.

The overall sequence of decrease, local increase, and final decrease in temperature therefore reflects the interplay of three competing mechanisms: advective and mixing-induced cooling, viscous dissipation (recovery heating), and conductive removal into the wall. At $t=4.6$\,s the throat surface reaches the ablation temperature and is thereafter held at 922\,K. With increasing time, viscous dissipation progressively elevates the near-wall gas temperature, steepening the local wall-normal gradient $(\partial T/\partial y)_w$. As a result, the instantaneous wall heat flux $q_w$ initially increases before gradually decaying as the system relaxes toward a new quasi-steady state.

Figure~\ref{fig:Temperature gradients} presents the temperature gradients at the nozzle throat for times between 3\,s and 6\,s. These trends are consistent with the corresponding static temperature profiles and the observed heat flux variation. A distinct sign change in the gradient curves indicates the presence of a local temperature overshoot near the wall. As this overshoot strengthens with time, the near-wall gradients become sharper, leading to a more negative slope. This behaviour reflects the dynamic balance between conduction into the cooled wall and viscous dissipation within the boundary layer.

At $t=3.0$\,s, the wall heat flux is highest owing to the large wall–fluid temperature difference. By $t=4.0$\,s the gradient magnitude decreases and the heat flux correspondingly weakens. However, at later times (5–6\,s) viscous heating remains significant and because the wall is held at a constant temperature of 922\,K during this interval, the near-wall overshoot is enhanced, restoring a steeper gradient. Consequently, the wall–fluid temperature contrast increases again, producing a rise in the instantaneous wall heat flux. Overall, the temporal evolution of the heat flux can be understood as the outcome of competing mechanisms: thermal conduction into the cold wall versus viscous heating within the flow.

\begin{figure}[hbt!]
\centering
\includegraphics[width=1\textwidth]{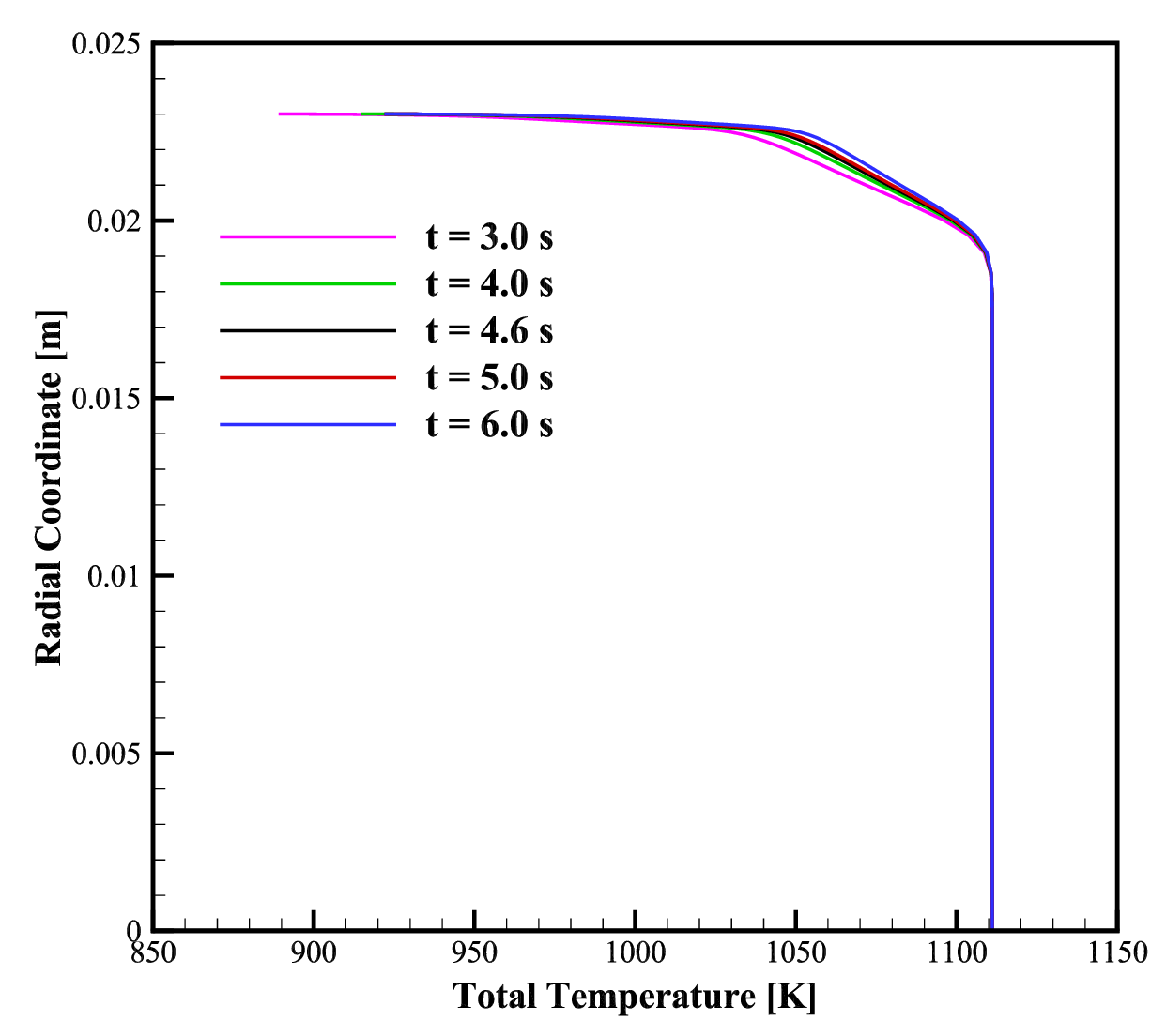}
\caption{Radial distribution of total temperature at the nozzle throat.}
\label{fig:throat_total_temperature}
\end{figure}

Figure ~\ref{fig:throat_total_temperature}  illustrates the total temperature distribution across the nozzle throat for times between 3\,s and 6\,s. In the core flow, the total temperature remains nearly uniform and constant, consistent with the freestream value. However, within the near-wall region, a gradual increase in total temperature is observed over time. At $t=3.0$\,s,  the cooled wall extracts thermal energy efficiently, resulting in a lower near-wall total temperature. At later times (4–6\,s), viscous dissipation within the boundary layer becomes more pronounced, leading to a slight but systematic rise in the total temperature near the wall. This behavior aligns with the static temperature overshoot and the steepening of temperature gradients previously discussed, confirming the competing influences of wall cooling and viscous heating in the throat region.

\subsection{Density and Velocity Profiles at the Throat}

\begin{figure}[hbt!]
    \centering
    \subfloat[Radial density profiles at the nozzle throat from 3 to 6 seconds.]{
        \includegraphics[width=0.48\textwidth]{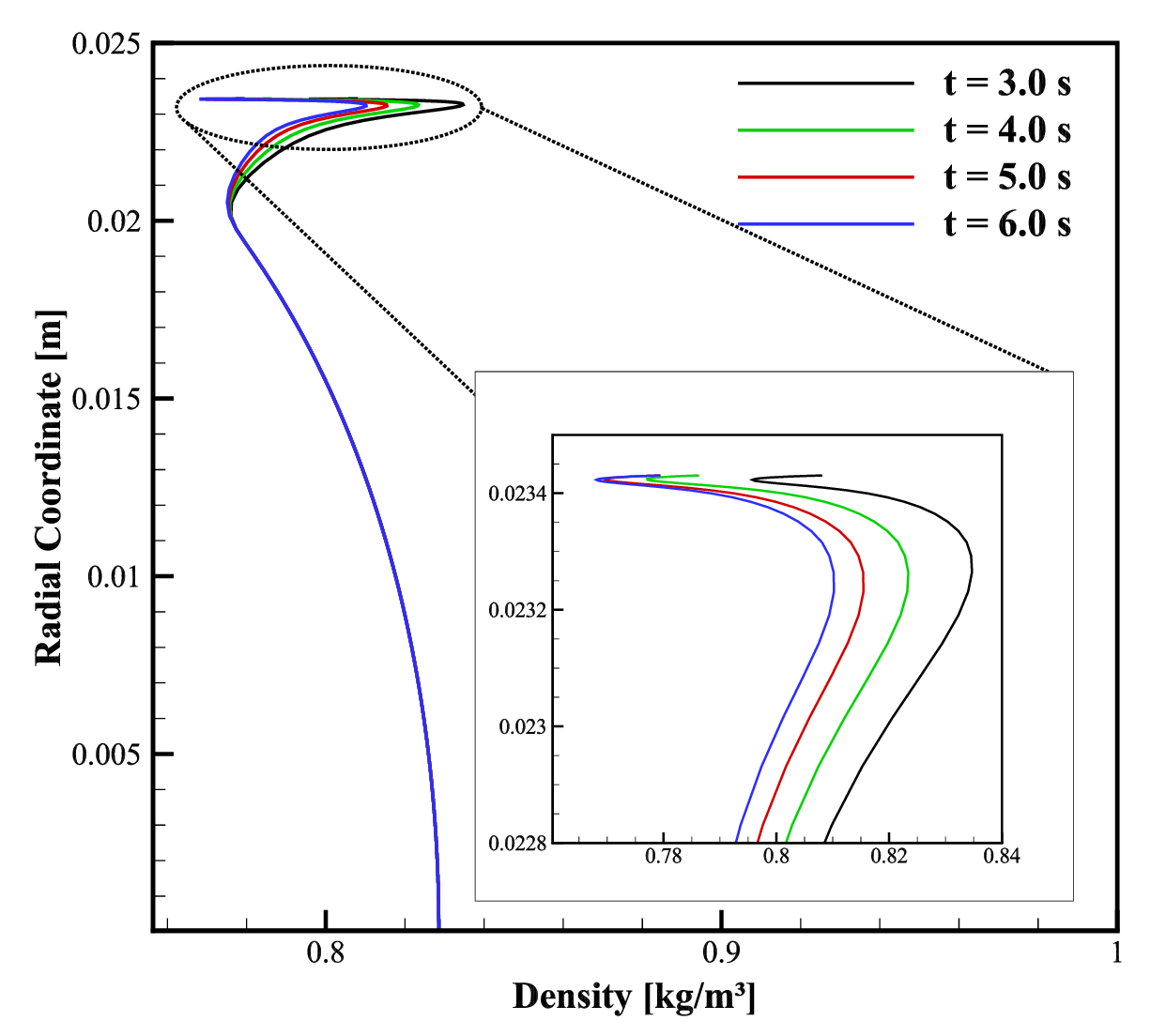}
        \label{fig:density_Throat_Profiles}
    }
    \hfill
    \subfloat[Axial velocity profiles at the nozzle throat from 3 to 6 seconds.]{
        \includegraphics[width=0.48\textwidth]{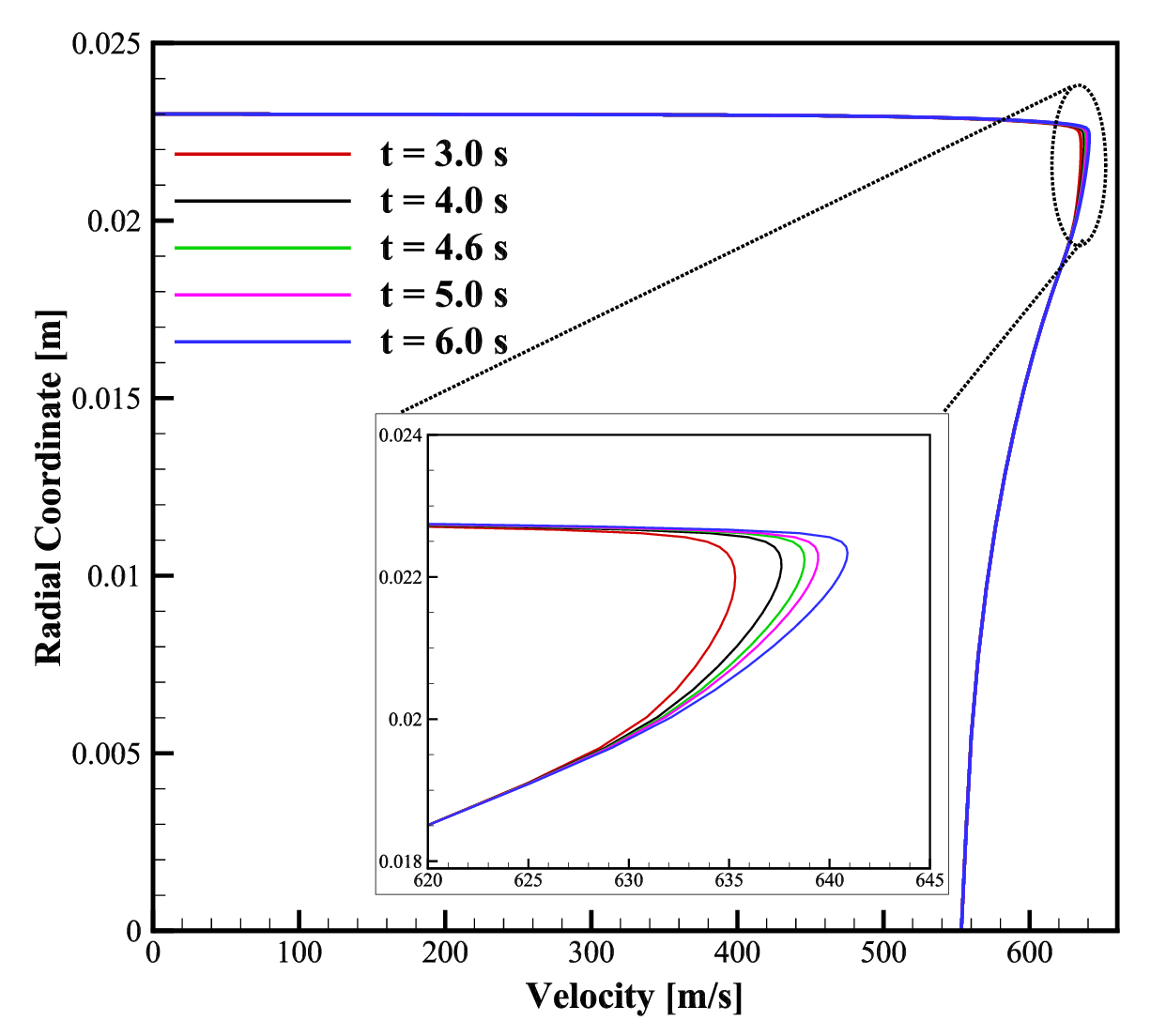}
        \label{fig:velocity_Throat_Profiles}
    }
    \caption{Density and axial velocity distributions across the nozzle throat at different times.}
    \label{fig:density and velocity}
\end{figure}

Since density is governed by the ideal gas law, it strongly depends on the local static temperature. From the density profile, as shown in Fig. ~\ref{fig:density_Throat_Profiles}, moving outward from the core flow toward the onset of the boundary layer, the density remains nearly constant. However, near the onset of the boundary layer, the temperature plot in Fig. ~\ref{fig:throat_static_temperature} shows a noticeable drop in static temperature, and accordingly the density increases in this region. Further into the boundary layer, viscous dissipation causes the static temperature to rise sharply, which leads to a local dip in density. Finally, at the cold wall, the density increases again in accordance with the temperature trend dictated by the ideal gas law.

The velocity profile in Fig.~\ref{fig:velocity_Throat_Profiles} shows that axial velocity increases in the same region where the density decreases, consistent with the formation of a low-density pocket driven by viscous heating. To satisfy local mass conservation, the velocity overshoots in this zone so that the local mass flux  is maintained, before decaying to zero at the wall due to the no-slip condition.

\subsection{Comparative Heat Flux Analysis Across Nozzle Regions}

To further elucidate the evolution of heat transfer in the TPS, three axial locations are examined: the throat ($x=0.091\ \mathrm{m}$), the converging section ($x=0.075\ \mathrm{m}$), and the diverging section ($x=0.105\ \mathrm{m}$). Results are presented for the first 60\,s of simulated time.

\begin{figure}[hbt!]
\centering
\includegraphics[width=1\textwidth]{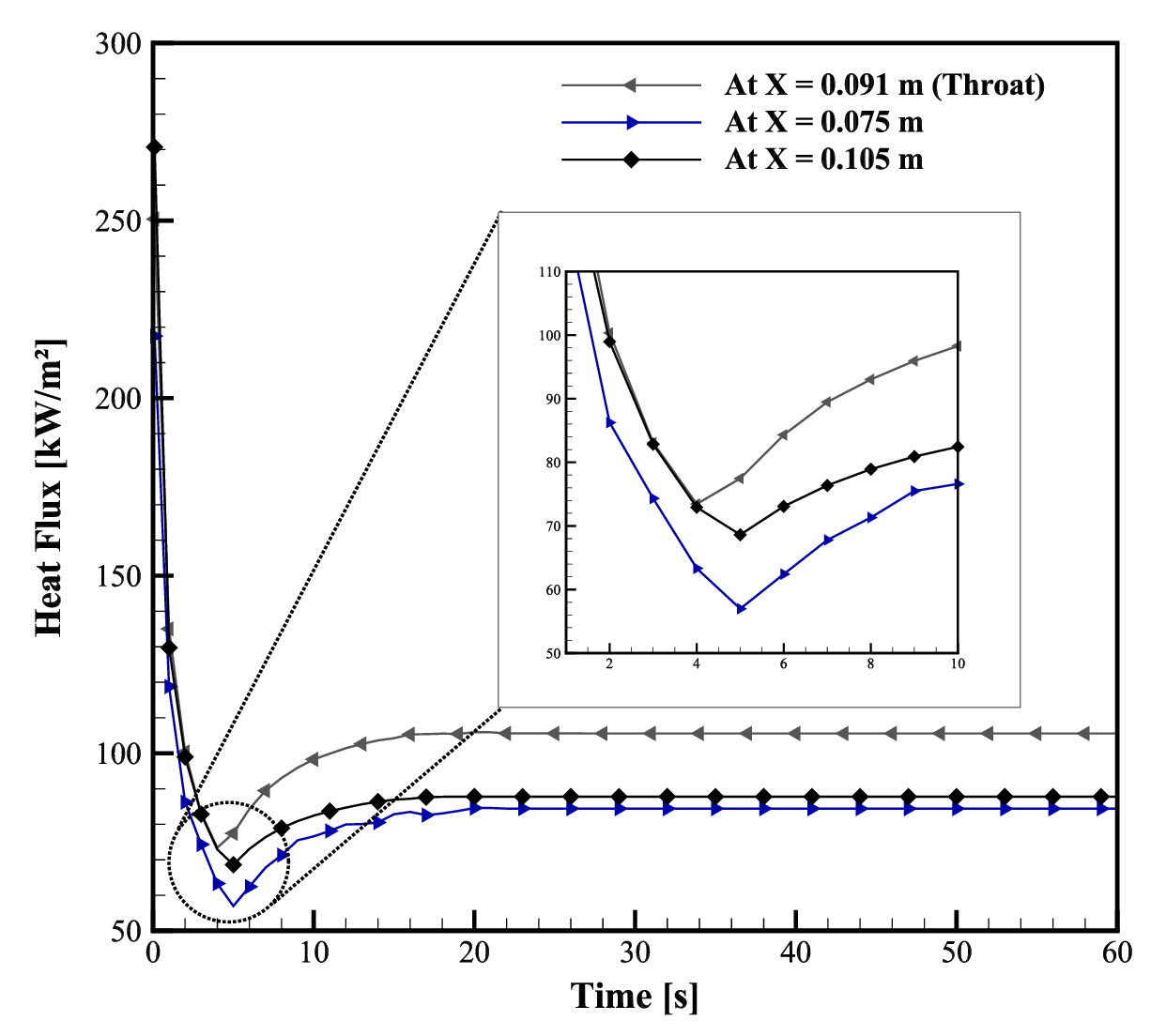}
\caption{Comparison of surface heat flux at three axial locations in the nozzle.}
\label{fig:Comparison Of Heat flux at three different X locations}
\end{figure}

The heat flux results are presented in Fig.~\ref{fig:Comparison Of Heat flux at three different X locations}. All three locations follow a similar trend, as discussed for the throat region. The maximum heat flux occurs in the throat region, followed by the diverging section, while the converging section experiences the least heat flux.

\begin{figure}[hbt!]
    \centering
    \subfloat[Throat region ($x=0.091\ \mathrm{m}$)]{
        \includegraphics[width=0.48\textwidth]{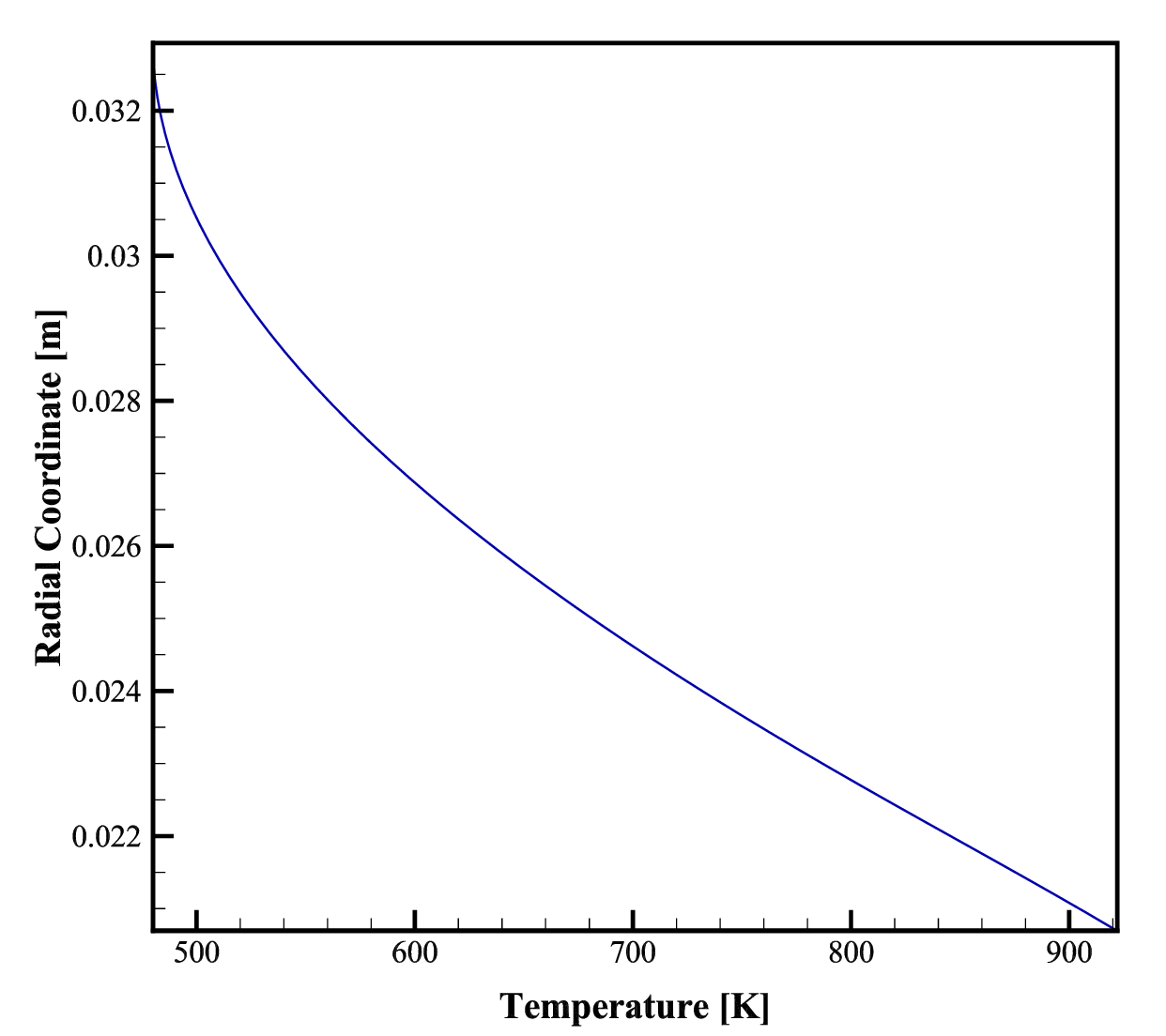}
        \label{fig:Temperature through the material at X= 0.091 m}
    }
    \hfill
    \subfloat[Converging section ($x=0.075\ \mathrm{m}$)]{
        \includegraphics[width=0.48\textwidth]{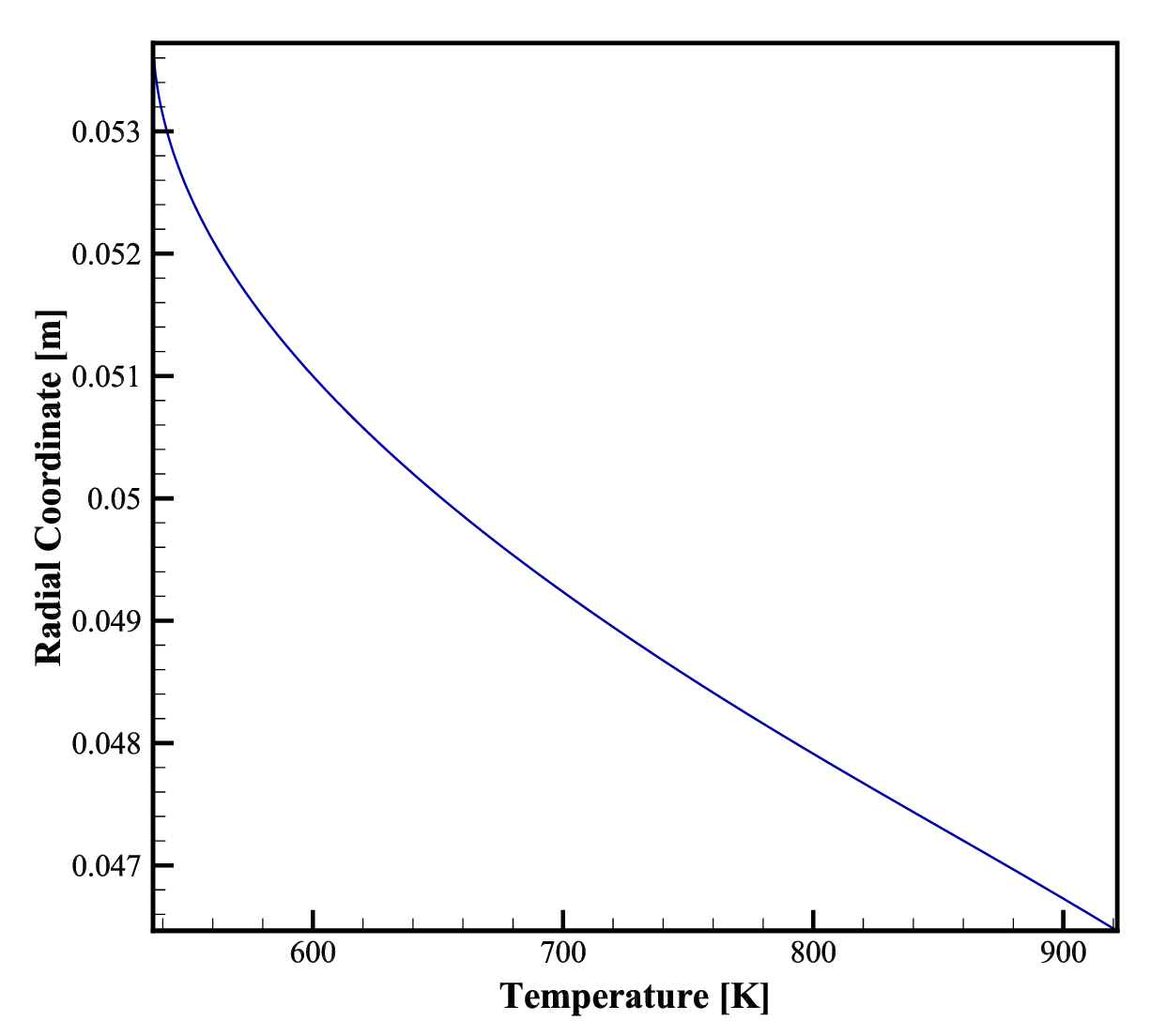}
        \label{fig:Temperature through the material at X= 0.075 m}
    }
    \caption{Temperature variation through the TPS thickness at the throat and converging section.}
\end{figure}

\begin{figure}[hbt!]
    \centering
    \subfloat[Diverging section ($x=0.105\ \mathrm{m}$)]{
        \includegraphics[width=0.48\textwidth]{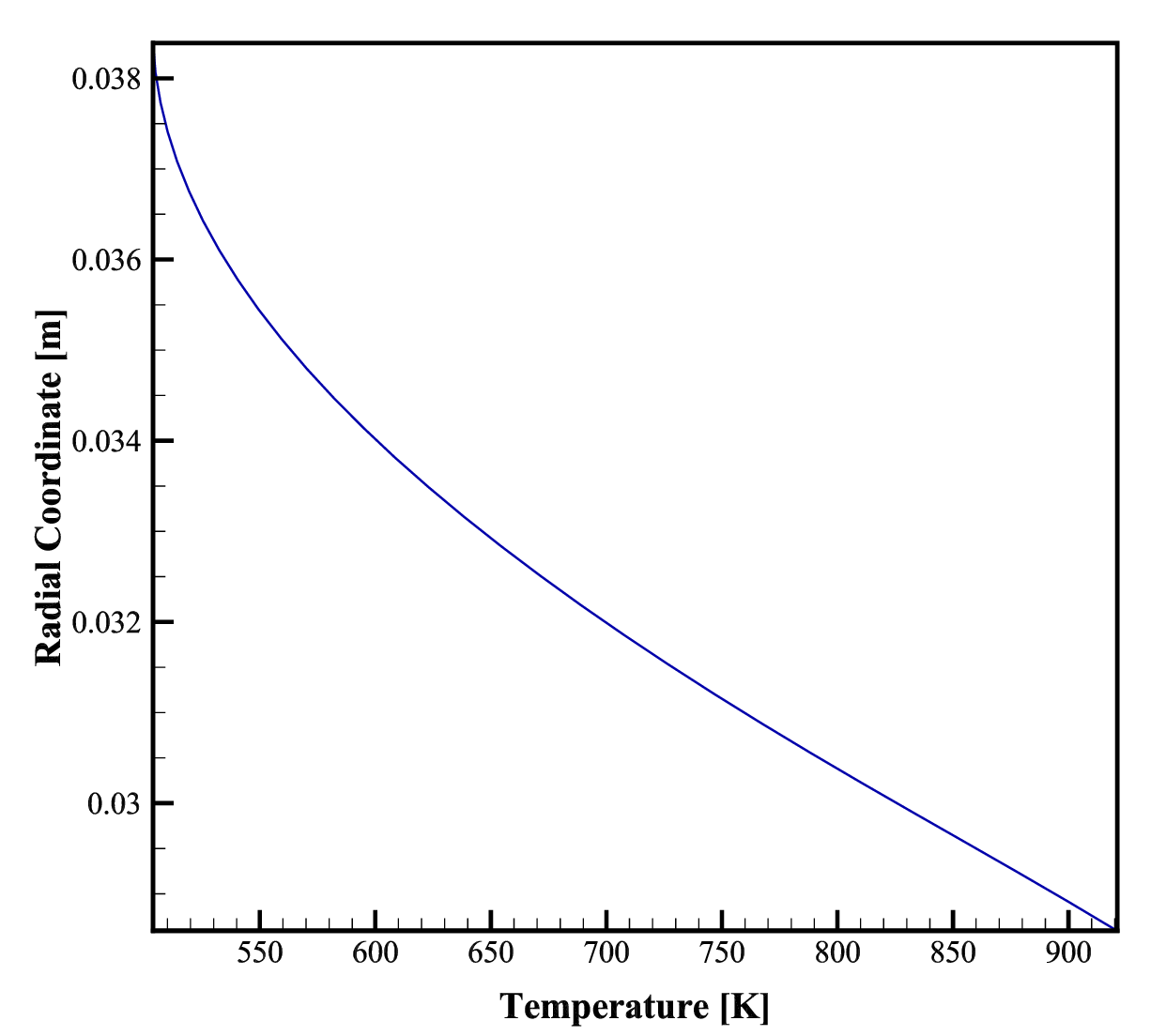}
        \label{fig:Temperature through the material at X= 0.105 m}
    }
    \hfill
    \subfloat[Temperature contour of the TPS material after 120\,s]{
        \includegraphics[width=0.48\textwidth]{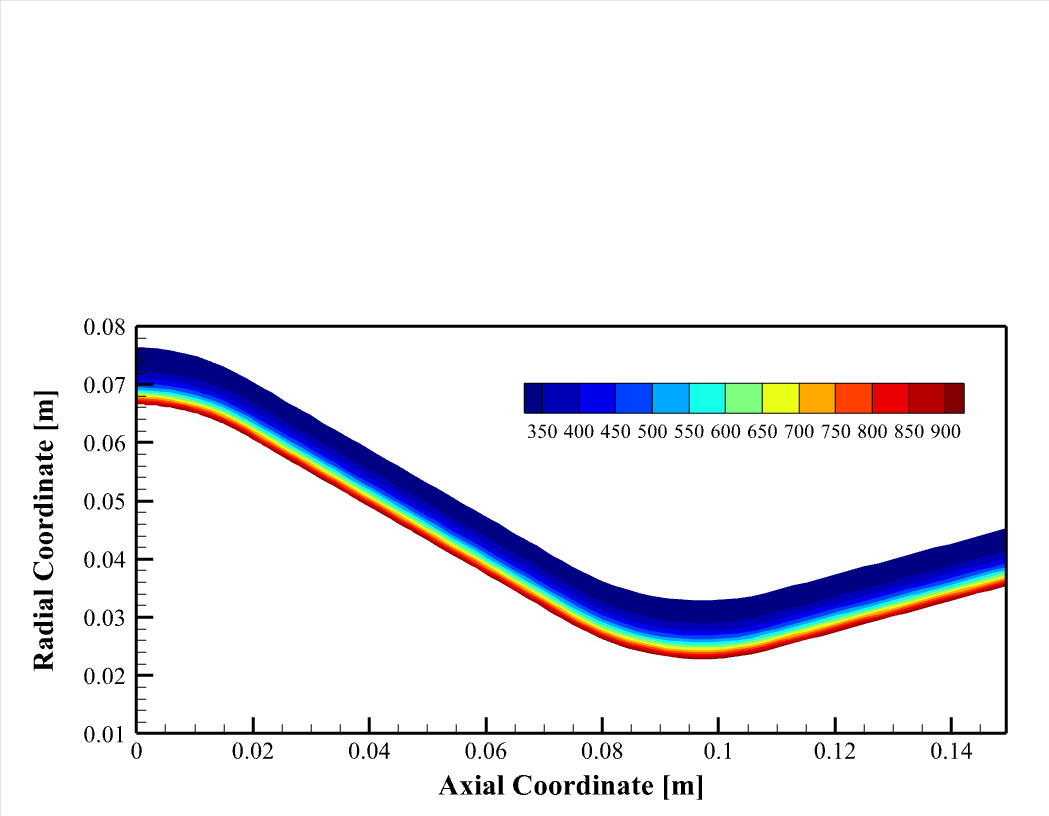}
        \label{fig:Temperature Profile of TPS}
    }
    \caption{Temperature variation at the diverging section and overall TPS temperature contour after 120\,s.}
\end{figure}

The temperature variation through the thickness of the TPS material at these locations is analyzed in Figs.~\ref{fig:Temperature through the material at X= 0.091 m}, \ref{fig:Temperature through the material at X= 0.075 m},  and \ref{fig:Temperature through the material at X= 0.105 m}. The surface exposed to the hot gas flow reaches a significantly higher temperature, while the back-face remains close to the adiabatic wall temperature. A gradual decrease in temperature through the TPS thickness is observed. In addition, the overall thermal response of the nozzle wall is illustrated by the temperature contour of the TPS material after 120\,s, shown in Fig.~\ref{fig:Temperature Profile of TPS}.

\subsection{Surface Recession}

Once the surface temperature of the nozzle TPS reaches the ablation temperature, the surface of the nozzle wall begins to erode due to surface recession. The surface recession over time is calculated by the Eq. \eqref{eq:surface recession}. The study of surface recession is critical as the dynamics of nozzle flow can notably get affected due to the surface ablation. 

\begin{figure}[hbt!]
    \centering
    \subfloat[Comparison of initial and receded nozzle wall geometries after 120 s of operation.]{
        \includegraphics[width=0.48\textwidth]{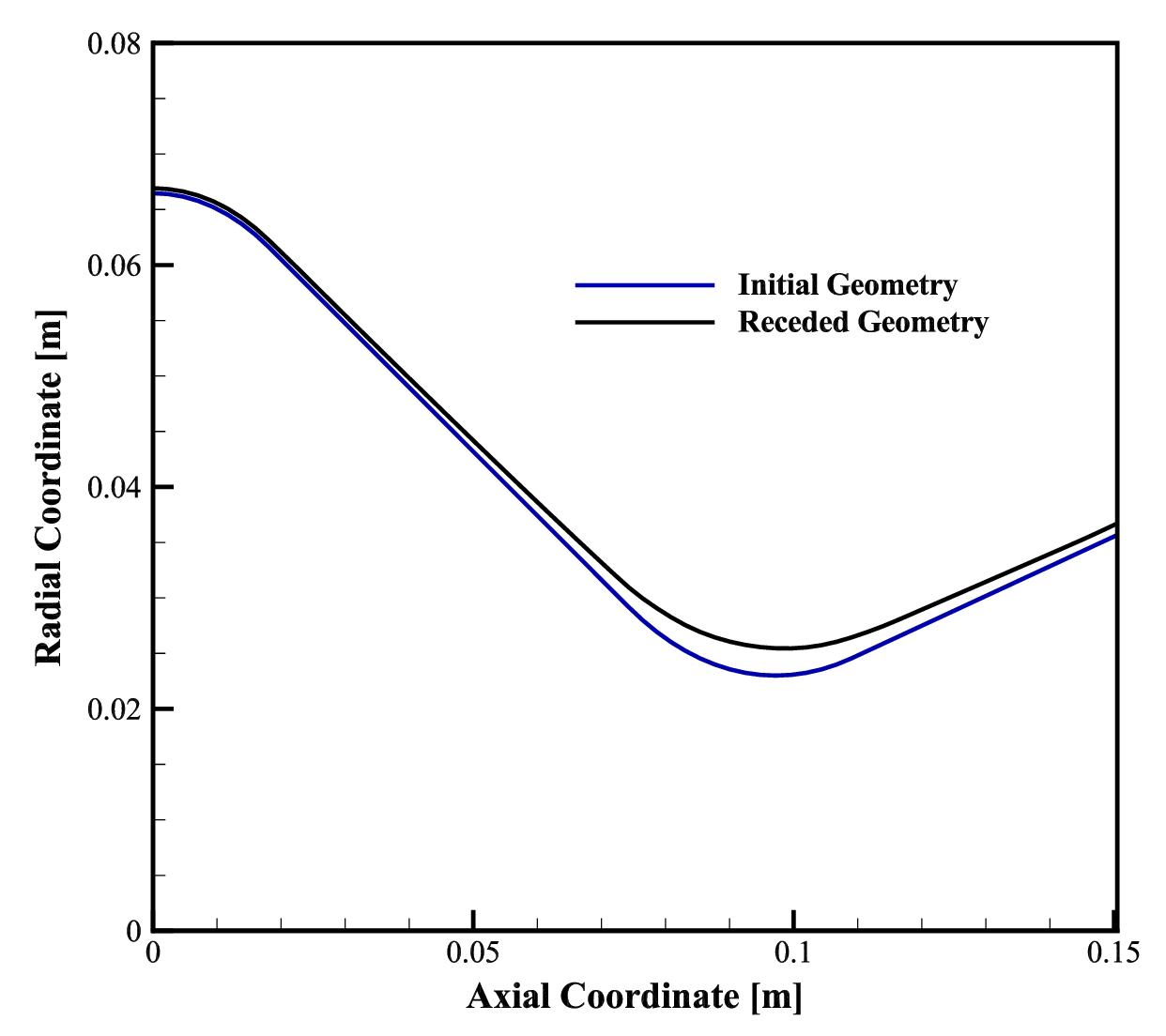}
        \label{fig:Initial Vs Receded Nozzle wall}
    }
    \hfill
    \subfloat[Axial distribution of surface recession along the nozzle wall.]{
        \includegraphics[width=0.48\textwidth]{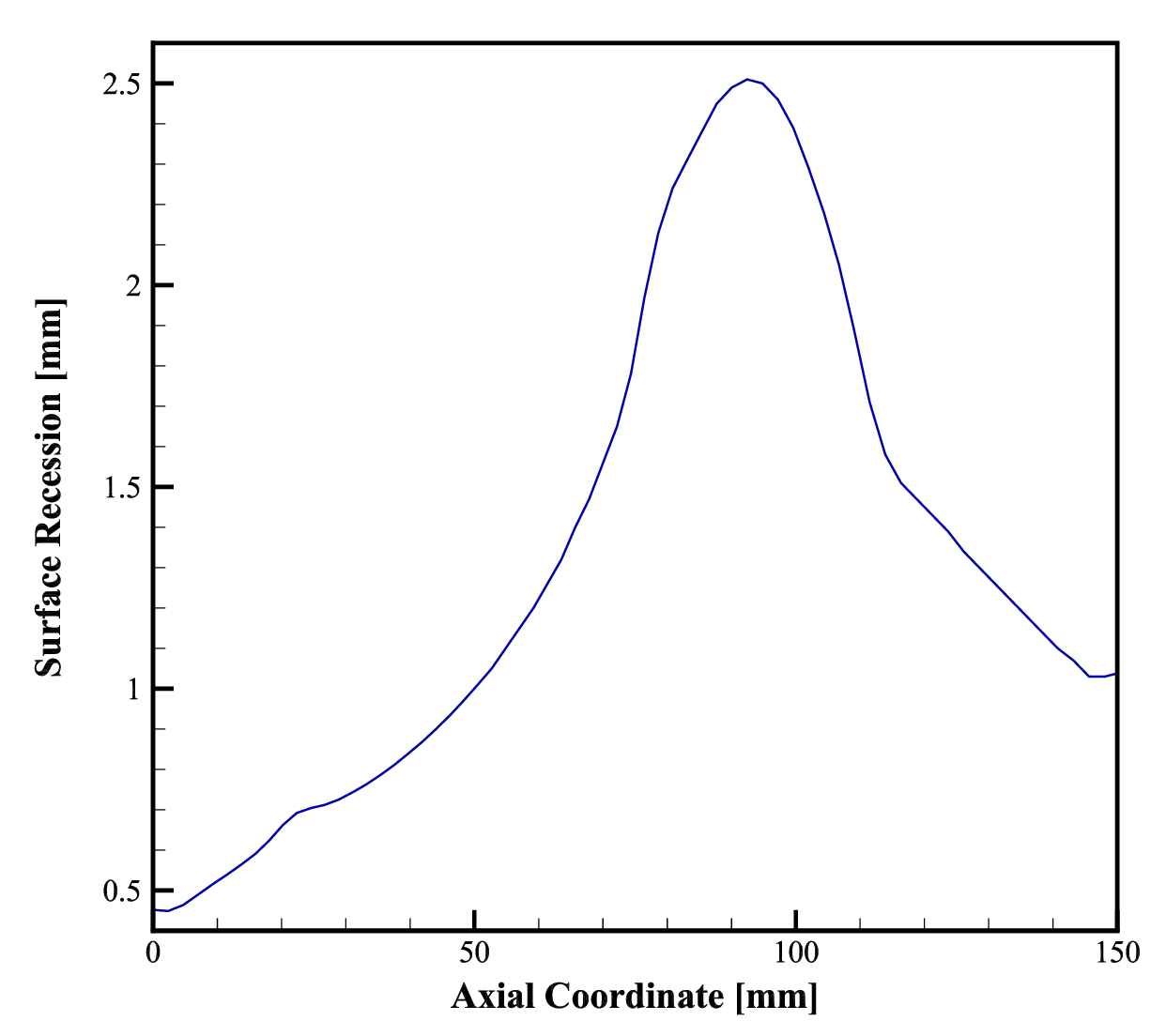}
        \label{fig:surface recession}
    }
    \caption{Surface recession.}
\end{figure}

The initial and the final nozzle wall geometries are shown in Fig.~\ref{fig:Initial Vs Receded Nozzle wall}. The receded geometry is plotted at 120 s of nozzle operation. It is evident that the maximum surface recession occurs at the throat of the nozzle compared to the other regions of the nozzle wall. The maximum recession at the nozzle throat is calculated to be approximately 2.5 mm, as can be visualized in Fig.~\ref{fig:surface recession}. The surface recession of TPS material aligns closely with the heat flux and surface temperature variations. 

\section{Conclusions}

A conjugate flow–thermal study of a convergent–divergent rocket nozzle protected by a charring ablative TPS (AVCOAT) was presented. The work coupled a commercial CFD solver (\textsc{ANSYS Fluent}) with an in-house transient material response code (CATS) using a weak, non-iterative exchange of boundary conditions at the fluid–solid interface. Pyrolysis-gas effects were accounted for via the Kays \cite{kays1980convective} blowing correction. 
The flow solver was validated against the NASA benchmark by Back \emph{et al.} (Case 262) \cite{back1964convective}, 
and the material solver was validated in prior studies \cite{appar2022conjugate,doi:10.1080/10618562.2021.2017900}. 
Together, these validations provide confidence in the coupled predictions.

The coupled simulations identified the nozzle throat as the most thermally critical location: the throat experiences the highest convective loading, attains the TPS ablation temperature earlier than adjacent regions, and exhibited the largest cumulative surface recession. Over 120\,s of simulated operation the maximum surface recession at the throat is approximately 2.5\,mm. Temporally, the wall heat flux exhibited an early peak, a transient reduction, and a later resurgence; this behaviour was explained by the competing effects of conduction to the cooled wall, viscous dissipation within the near-wall flow, and the evolving surface temperature as ablation and charring progressed. The model captured interacting mechanisms including pyrolysis, blowing corrections, viscous heating and conduction, thereby providing a predictive framework for TPS response under extreme nozzle heat loads.
Overall, the presented framework offers a practical and physically consistent approach to predict AVCOAT TPS behaviour under extreme nozzle environments and highlights throat regions as priority locations for enhanced thermal protection and further study.

\section*{Acknowledgments}
The authors acknowledge the financial support provided by the Indian Space Research Organization (ISRO) through the Grant No. STC/AE/2024427L. The authors also thank the National Supercomputing Mission (NSM) for providing computing resources of “PARAM Sanganak” at IIT Kanpur, which is implemented by C-DAC and supported by the Ministry of Electronics and Information Technology (MeitY) and the Department of Science and Technology (DST), Government of India. 

\section*{Author Declarations}
%\subsection*{Conflict of Interest}
The authors have no conflict of interest to disclose.
	
\section*{Data Availability}
	
The data that support the ﬁndings of this study can be made available by the corresponding author upon a reasonable request.

\clearpage

\bibliography{references}

\begin{thebibliography}{36}
\newcommand{\enquote}[1]{``#1''}
\providecommand{\natexlab}[1]{#1}
\providecommand{\url}[1]{\texttt{#1}}
\providecommand{\urlprefix}{URL }
\expandafter\ifx\csname urlstyle\endcsname\relax
  \providecommand{\doi}[1]{\discretionary{}{}{}https://doi.org/#1}\else
  \providecommand{\doi}[1]{\discretionary{}{}{}\urlstyle{rm}\url{https://doi.org/#1}}\fi

\bibitem[{Powers et~al.(1981)Powers, Bailey, and Morrison}]{powers1981shuttle}
Powers, L., Bailey, R., and Morrison, B., \enquote{Shuttle solid rocket motor nozzle alternate ablative evaluation,} \emph{17th Joint Propulsion Conference}, 1981, p. 1461.
\newblock \doi{10.2514/6.1981-1461}.

\bibitem[{Arnold et~al.(1979)Arnold, Dodson, and Laub}]{arnold1979subscale}
Arnold, J., Dodson, J., and Laub, B., \enquote{Subscale solid motor nozzle tests, phase 4 and nozzle materials screening and thermal characterization, phase 5,} Tech. rep., 1979.

\bibitem[{Darbandi and Roohi(2011)}]{darbandi2011study}
Darbandi, M., and Roohi, E., \enquote{Study of subsonic--supersonic gas flow through micro/nanoscale nozzles using unstructured DSMC solver,} \emph{Microfluidics and nanofluidics}, Vol.~10, 2011, pp. 321--335.
\newblock \doi{10.1007/s10404-010-0671-7}.

\bibitem[{Stark(2013)}]{stark2013flow}
Stark, R.~H., \enquote{Flow separation in rocket nozzles-an overview,} \emph{49th AIAA/ASME/SAE/ASEE Joint PropulsionConference}, 2013, p. 3840.
\newblock \doi{10.2514/6.2013-3840}.

\bibitem[{Rakhsha et~al.(2023)Rakhsha, Zargarabadi, and Saedodin}]{rakhsha2023effect}
Rakhsha, S., Zargarabadi, M.~R., and Saedodin, S., \enquote{The effect of nozzle geometry on the flow and heat transfer of pulsed impinging jet on the concave surface,} \emph{International Journal of Thermal Sciences}, Vol. 184, 2023, p. 107925.
\newblock \doi{10.1016/j.ijthermalsci.2022.107925}.

\bibitem[{Lijo et~al.(2010)Lijo, Kim, Setoguchi, and Matsuo}]{lijo2010numerical}
Lijo, V., Kim, H.~D., Setoguchi, T., and Matsuo, S., \enquote{Numerical simulation of transient flows in a rocket propulsion nozzle,} \emph{International Journal of Heat and Fluid Flow}, Vol.~31, No.~3, 2010, pp. 409--417.
\newblock \doi{10.1016/j.ijheatfluidflow.2009.12.005}.

\bibitem[{Schneider et~al.(2018)Schneider, G{\'e}nin, Stark, Oschwald, Karl, and Hannemann}]{schneider2018numerical}
Schneider, D., G{\'e}nin, C., Stark, R., Oschwald, M., Karl, S., and Hannemann, V., \enquote{Numerical model for nozzle flow application under liquid oxygen/methane hot-flow conditions,} \emph{Journal of Propulsion and Power}, Vol.~34, No.~1, 2018, pp. 221--233.
\newblock \doi{10.2514/1.B36611}.

\bibitem[{Balabel et~al.(2011)Balabel, Hegab, Nasr, and El-Behery}]{balabel2011assessment}
Balabel, A., Hegab, A., Nasr, M., and El-Behery, S.~M., \enquote{Assessment of turbulence modeling for gas flow in two-dimensional convergent--divergent rocket nozzle,} \emph{Applied Mathematical Modelling}, Vol.~35, No.~7, 2011, pp. 3408--3422.
\newblock \doi{10.1016/j.apm.2011.01.013}.

\bibitem[{Sathish(2017)}]{sathish2017heat}
Sathish, T., \enquote{Heat Transfer Analysis Of Nano-fluid Flow In A Converging Nozzle With Different Aspect Ratios.} \emph{Journal of New Materials for Electrochemical Systems}, Vol.~20, No.~4, 2017.
\newblock \doi{10.14447/jnmes.v20i4.321}.

\bibitem[{Thongsri et~al.(2022)Thongsri, Srathonghuam, and Boonpan}]{thongsri2022gas}
Thongsri, J., Srathonghuam, K., and Boonpan, A., \enquote{Gas flow and ablation of 122 mm supersonic rocket nozzle investigated by conjugate heat transfer analysis,} \emph{Processes}, Vol.~10, No.~9, 2022, p. 1823.
\newblock \doi{10.3390/pr10091823}.

\bibitem[{Sae-ngow et~al.(2021)Sae-ngow, Palsarn, and Boonpan}]{sae2021insulation}
Sae-ngow, C., Palsarn, S., and Boonpan, A., \enquote{Insulation analysis for rocket’s nozzle to reduce deformation of nozzle shape,} \emph{35th ME-NETT}, 2021, pp. 20--23.

\bibitem[{Hui et~al.(2017)Hui, Bao, Wei, and Liu}]{hui2017ablation}
Hui, W.-h., Bao, F.-t., Wei, X.-g., and Liu, Y., \enquote{Ablation performance of a 4D-braided C/C composite in a parameter-variable channel of a Laval nozzle in a solid rocket motor,} \emph{New Carbon Materials}, Vol.~32, No.~4, 2017, pp. 365--373.
\newblock \doi{10.1016/S1872-5805(17)60128-8}.

\bibitem[{Cross and Boyd(2018)}]{cross2018reduced}
Cross, P.~G., and Boyd, I.~D., \enquote{Reduced reaction mechanism for rocket nozzle ablation simulations,} \emph{Journal of Thermophysics and Heat Transfer}, Vol.~32, No.~2, 2018, pp. 429--439.

\bibitem[{Zhang et~al.(2022)Zhang, Wang, Wang, Lu, Yu, and Tian}]{zhang2022numerical}
Zhang, X., Wang, Z., Wang, R., Lu, C., Yu, R., and Tian, H., \enquote{Numerical simulation of chemical ablation and mechanical erosion in hybrid rocket nozzle,} \emph{Acta Astronautica}, Vol. 192, 2022, pp. 82--96.
\newblock \doi{10.1016/j.actaastro.2021.12.012}.

\bibitem[{Babu and Murthy(2020)}]{babu2020prediction}
Babu, G.~V., and Murthy, V.~B., \enquote{Prediction of thermal ablation in rocket nozzle using CFD and FEA,} \emph{International Journal of Computational Materials Science and Engineering}, Vol.~9, No.~03, 2020, p. 2050014.
\newblock \doi{10.1142/S2047684120500141}.

\bibitem[{Bartz(1957)}]{bartz1957simple}
Bartz, D.~R., \enquote{A simple equation for rapid estimation of rocket nozzle convective heat transfer coefficients,} \emph{Jet Propul.}, Vol.~27, 1957, pp. 49--51.

\bibitem[{Cross and Boyd(2017)}]{cross2017conjugate}
Cross, P.~G., and Boyd, I.~D., \enquote{Conjugate analysis of rocket nozzle ablation,} \emph{47th AIAA Thermophysics Conference}, 2017, p. 3351.
\newblock \doi{10.2514/6.2017-3351}.

\bibitem[{Cross and Boyd(2019)}]{cross2019conjugate}
Cross, P.~G., and Boyd, I.~D., \enquote{Conjugate analyses of ablation in rocket nozzles,} \emph{Journal of Spacecraft and Rockets}, Vol.~56, No.~5, 2019, pp. 1593--1610.

\bibitem[{Zhang(2011)}]{zhang2011coupled}
Zhang, X., \enquote{Coupled simulation of heat transfer and temperature of the composite rocket nozzle wall,} \emph{Aerospace Science and Technology}, Vol.~15, No.~5, 2011, pp. 402--408.
\newblock \doi{10.1016/j.ast.2010.09.006}.

\bibitem[{Ding et~al.(2017)Ding, Wang, and Wang}]{ding2017transient}
Ding, H., Wang, C., and Wang, G., \enquote{Transient conjugate heat transfer in critical flow nozzles,} \emph{International Journal of Heat and Mass Transfer}, Vol. 104, 2017, pp. 930--942.
\newblock \doi{10.1016/j.ijheatmasstransfer.2016.09.021}.

\bibitem[{Pizzarelli et~al.(2016)Pizzarelli, Nasuti, Votta, and Battista}]{pizzarelli2016validation}
Pizzarelli, M., Nasuti, F., Votta, R., and Battista, F., \enquote{Validation of conjugate heat transfer model for rocket cooling with supercritical methane,} \emph{Journal of Propulsion and Power}, Vol.~32, No.~3, 2016, pp. 726--733.
\newblock \doi{10.2514/1.B35945}.

\bibitem[{Guan et~al.(2017)Guan, Zhang, and Shan}]{guan2017conjugate}
Guan, T., Zhang, J.-z., and Shan, Y., \enquote{Conjugate heat transfer on leading edge of a conical wall subjected to external cold flow and internal hot jet impingement from chevron nozzle--Part 2: Numerical analysis,} \emph{International Journal of Heat and Mass Transfer}, Vol. 106, 2017, pp. 339--355.
\newblock \doi{10.1016/j.ijheatmasstransfer.2016.10.048}.

\bibitem[{Kuntz et~al.(2001)Kuntz, Hassan, and Potter}]{kuntz2001predictions}
Kuntz, D.~W., Hassan, B., and Potter, D.~L., \enquote{Predictions of ablating hypersonic vehicles using an iterative coupled fluid/thermal approach,} \emph{Journal of thermophysics and Heat Transfer}, Vol.~15, No.~2, 2001, pp. 129--139.
\newblock \doi{10.2514/2.6594}.

\bibitem[{Olynick et~al.(1999)Olynick, Chen, and Tauber}]{olynick1999aerothermodynamics}
Olynick, D., Chen, Y.-K., and Tauber, M.~E., \enquote{Aerothermodynamics of the Stardust sample return capsule,} \emph{Journal of Spacecraft and Rockets}, Vol.~36, No.~3, 1999, pp. 442--462.

\bibitem[{Thompson and Gnoffo(2008)}]{thompson2008implementation}
Thompson, R., and Gnoffo, P., \enquote{Implementation of a blowing boundary condition in the LAURA code,} \emph{46th AIAA Aerospace Sciences Meeting and Exhibit}, 2008, p. 1243.
\newblock \doi{10.2514/6.2008-1243}.

\bibitem[{Wiebenga and Boyd(2012)}]{wiebenga2012computation}
Wiebenga, J., and Boyd, I., \enquote{Computation of multi-dimensional material response coupled to hypersonic flow,} \emph{43rd AIAA Thermophysics Conference}, 2012, p. 2873.
\newblock \doi{10.2514/6.2012-2873}.

\bibitem[{Alkandry et~al.(2013)Alkandry, Boyd, and Martin}]{alkandry2013coupled}
Alkandry, H., Boyd, I.~D., and Martin, A., \enquote{Coupled flow field simulations of charring ablators with nonequilibrium surface chemistry,} \emph{44th AIAA Thermophysics Conference}, 2013, p. 2634.
\newblock \doi{10.2514/6.2013-2634}.

\bibitem[{Martin and Boyd(2015)}]{doi:10.2514/1.A32847}
Martin, A., and Boyd, I.~D., \enquote{Strongly Coupled Computation of Material Response and Nonequilibrium Flow for Hypersonic Ablation,} \emph{Journal of Spacecraft and Rockets}, Vol.~52, No.~1, 2015, pp. 89--104.
\newblock \doi{10.2514/1.A32847}.

\bibitem[{Thakre and Yang(2008)}]{thakre2008chemical}
Thakre, P., and Yang, V., \enquote{Chemical erosion of carbon-carbon/graphite nozzles in solid-propellant rocket motors,} \emph{Journal of Propulsion and Power}, Vol.~24, No.~4, 2008, pp. 822--833.
\newblock \doi{10.2514/1.34946}.

\bibitem[{Back et~al.(1964)Back, Massier, and Gier}]{back1964convective}
Back, L., Massier, P., and Gier, H., \enquote{Convective heat transfer in a convergent-divergent nozzle,} \emph{International Journal of Heat and Mass Transfer}, Vol.~7, No.~5, 1964, pp. 549--568.
\newblock \doi{10.1016/0017-9310(64)90052-3}.

\bibitem[{Manual(2009)}]{manual2009ansys}
Manual, U., \enquote{ANSYS FLUENT 12.0,} \emph{Theory Guide}, Vol.~67, 2009.

\bibitem[{Appar et~al.(2022)Appar, Kumar, and Naspoori}]{appar2022conjugate}
Appar, A., Kumar, R., and Naspoori, S.~K., \enquote{Conjugate flow-thermal analysis of a hypersonic reentry vehicle in the rarefied flow regime,} \emph{Physics of Fluids}, Vol.~34, No.~2, 2022.
\newblock \doi{10.1063/5.0082783}.

\bibitem[{Kays et~al.(1980)Kays, Crawford, and Weigand}]{kays1980convective}
Kays, W.~M., Crawford, M.~E., and Weigand, B., \emph{Convective heat and mass transfer}, Vol.~4, McGraw-Hill New York, 1980.

\bibitem[{Phadnis et~al.(2020)Phadnis, Raveendranath, and Jayachandran}]{phadnis2020effect}
Phadnis, T.~R., Raveendranath, P., and Jayachandran, T., \enquote{Effect of ply orientation on the in-depth response of carbon-phenolic ablative,} \emph{Journal of Thermophysics and Heat Transfer}, Vol.~34, No.~3, 2020, pp. 650--658.
\newblock \doi{10.2514/1.T5877}.

\bibitem[{Appar and Kumar(2021)}]{doi:10.1080/10618562.2021.2017900}
Appar, A., and Kumar, R., \enquote{Effect of Thermal Ablation at the Fluid-Solid Interface of a Hypersonic Reentry Vehicle in Rarefied Flow Regime,} \emph{International Journal of Computational Fluid Dynamics}, Vol.~35, No.~8, 2021, pp. 610--631.
\newblock \doi{10.1080/10618562.2021.2017900}.

\bibitem[{Williams(1992)}]{williams1992thermal}
Williams, S., \emph{Thermal protection materials: thermophysical property data}, Vol. 1289, National Aeronautics and Space Administration, Office of Management, 1992.

\end{thebibliography}

\end{document}